\documentclass[a4paper,11pt]{article}
\pdfoutput=1 

\usepackage{jheppub} 
\usepackage[T1]{fontenc} 
\usepackage{slashed}
\usepackage{arydshln}
\usepackage{amsmath}
\usepackage{amssymb}
\usepackage{amsfonts}
\usepackage{epsfig,multicol,bbm}
\newcommand{\be}{\begin{equation}}
\newcommand{\ee}{\end{equation}}
\newcommand{\bea}{\begin{eqnarray}}
\newcommand{\eea}{\end{eqnarray}}

\newcommand{\Eqs}[2]{Eqs.~\eqref{#1} and \eqref{#2}}

\def\nn{\nonumber}

\title{\boldmath Dark gauge boson production from neutron stars via nucleon-nucleon bremsstrahlung}

\author[a,b]{Chang Sub Shin}
\author[c,d]{, Seokhoon Yun}

\affiliation[a]{Center for Theoretical Physics of the Universe, \\ Institute for Basic Science (IBS), Daejeon, 34126, Korea}
\affiliation[b]{Department of Physics and Institute of Quantum Systems (IQS), \\ Chungnam National University, Daejeon 34134, Republic of Korea}
\affiliation[c]{Dipartimento di Fisica e Astronomia, Universit\`a degli Studi di Padova, \\ Via Marzolo 8, 35131 Padova, Italy}
\affiliation[d]{Istituto Nazionale di Fisica Nucleare (INFN), Sezione di Padova, \\ Via Marzolo 8, 35131 Padova, Italy}

\emailAdd{csshin@cnu.ac.kr}
\emailAdd{seokhoon.yun@pd.infn.it}

\abstract{We discuss the dark gauge boson emission from neutron stars via nucleon-nucleon bremsstrahlung. Through the rigorous treatment of the effective field theory prescription and the thermal effect, we derive the relevant couplings of dark gauge bosons to hadrons in medium. As a specific example, the $U(1)_{\rm B-L}$ gauge boson scenario is chosen to investigate dark gauge boson emissivities during supernovae and cooling of young neutron stars. From  the stellar cooling argument, we obtain the constraints on the $\rm B-L$ gauge coupling for given gauge boson masses  in two observations: the duration of the supernova neutrino signal of SN1987A, and the inferred x-ray luminosity of the compact object in the remnant of SN1987A (NS1987A). 
 In particular, the constraint from  SN1987A on the $U(1)_{\rm B-L}$ gauge boson scenario is revisited. The excluded gauge coupling due to the emission of transverse polarizations is an order of magnitude enhanced compared to the previous derivation. There is also a newly excluded parameter space due to the emission of longitudinal polarizations.
}

\begin{document} 
\maketitle
\flushbottom


\section{Introduction}
\label{sec:intro}


A number of observational conundrums in particle physics such as dark matter implications and baryon asymmetry point out an inevitable extension of the Standard Model (SM).
Many theoretical models for physics beyond the SM forecast a set of hypothetical particles, which are secluded from observations and known as `dark' sectors.
In particular, additional new $U(1)$ gauge bosons are ubiquitous in such models, e.g. string compactification as the unified theory of all forces.
These new gauge bosons become a mediator for interactions among hidden sector particles themselves, but also can be involved in connection between the hidden sector and the visible sector as the concept of `portal'.
If new $U(1)$ gauge bosons couple to the SM particles feebly, its mass could be lighter than the electroweak scale.
We call such a light weakly interacting gauge boson a dark gauge boson dubbed as $A^\prime_\mu$.

Stellar objects are good laboratories~\cite{Raffelt:1996wa} to examine such new particles in the hidden sector, including dark gauge bosons.
One of the main streams in the methodology of probing a new particle signal from stellar objects is the stellar cooling argument: when there is an extra energy leakage inside stars via free-streaming of novel particles produced, it could modify a thermal evolution of stars, e.g. the brightness at the tip of red giants or lifetime of horizontal branch stars.
If the observations are well fitted by simulations based on the SM, one can obtain the constraints on parametric space of hidden sector particles.

Neutrons stars (NS) are a typical remnant of supernova explosions, and an appealing stellar object to deduce physics at the fundamental level due to their interior under such an extreme circumstance.
There are several observations associated with NS, to which the stellar cooling argument can be applied.
A stringent constraint emerges from the measurement of the neutrino flux of SN1987A~\cite{Kamiokande-II:1987idp,Bionta:1987qt,Alekseev:1987ej} : the time duration of the bulk neutrino flux, which is fitted by the simulations~\cite{Mayle:1987as,Mayle:1989yx,Burrows:1988ah}, could be reduced by an extra cooling.
Taking into account the well established criterion for SN1987A that novel particle emissions cannot go over the limit of $3 \times 10^{52}\,{\rm erg} \, {\rm s}^{-1}$~\cite{Raffelt:1996wa}, one can derive the strong constraint on new particles such as axions~\cite{Turner:1987by,Raffelt:1987yt,Fischer:2016cyd,Chang:2018rso,Carenza:2019pxu,Croon:2020lrf,Fischer:2021jfm,Caputo:2021rux,Choi:2021ign} and dark gauge bosons~\cite{Dent:2012mx,An:2014twa,Chang:2016ntp,Hardy:2016kme,Knapen:2017xzo}.
Furthermore, young NS with ages under $10^4$-$10^5$ yr are also of interest in terms of the stellar cooling argument since they still possess a vigorous thermodynamic energy inside with a little contamination from any heating sources if it is well isolated.
The two observations reveal the possible presence of young isolated NS as a compact remnant of the supernova explosion: the predicted location~\cite{Cigan:2019shp} of NS1987A with its luminosity~\cite{Page:2020gsx}, and the x-ray spectrum of Cas A~\cite{1999IAUC7246,Hughes:1999ph}.
The observed data of their spectral luminosity match well with the cooling simulations based on the standard scenario~\cite{Page:2004fy,Yakovlev:2004iq,Page:2009fu}, in particular the rapid cooling rate of Cas A~\cite{Heinke:2010cr,Posselt:2018xaf,Wijngaarden:2019tht,Ho:2019vbn} implies the direct evidence of the superfluid phase~\cite{Page:2010aw,Shternin:2010qi,Ho:2014pta} in NS.
Hence, the stellar cooling argument are viable with early cooling histories of NS, and also applied to several light hidden particles such as axions~\cite{Leinson:2014ioa,Hamaguchi:2018oqw,Leinson:2021ety} and dark gauge bosons~\cite{Hong:2020bxo,Lu:2021uec}.

In this paper, we discuss the dark gauge boson production inside NS and its implications in terms of the stellar cooling argument.
It is known that nucleon-nucleon bremsstrahlung through exchanges of mediators associated with the strong interaction typically gives a dominant contribution to the production of novel particles due to kinematics with the quantum effect such as a degeneracy.\footnote{There are some exceptions that such a kinematic interruption can be resolved. For example, in the presence of nucleon superfluidity, the band structure with a gap in the dispersion relation of nucleons opens up a pseudo-decay channel, which is so called the pair breaking and formation process.}
To compute the rate, we need to specify the couplings of dark gauge bosons to SM particles, and in particular to hadrons.
Despite of the non-perturbative strong coupling regime, we have the prescription to determine the couplings in the effective lagrangian within the Chiral Perturbation Theory framework~\cite{Weinberg:1978kz,Gasser:1983yg,Gasser:1984gg,Pich:1995bw}; matching currents with the same behavior under a chiral transformation.
Apart from the manifest dark gauge boson couplings to the two-point currents of hadrons (e.g., $\bar{N}\gamma^\mu N$ for a nucleon field $N$), there could be interactions to the three-point currents~\cite{Kroll:1953vq} such as $A^\prime_\mu \pi^+ \bar{p}\gamma^\mu \gamma^5 n$, which are not considered in the literature.
Moreover, we will consider the thermal field theory~\cite{Laine:2016hma,Bellac:2011kqa} to account for the effective couplings in a medium~\cite{Hardy:2016kme}.
Interestingly, in the dark gauge boson scenario with the $\rm B-L$  (the baryon number minus the lepton number) charge assignment,  the dark gauge boson interaction to protons is screened by such a medium effect~\cite{Hong:2020bxo}. This gives rise to a net difference in the effective couplings to nucleons, whose effect is not taken into account carefully in the literature. As we will discuss, such a difference is one of the key ingredients for constraining the $U(1)_{\rm B-L}$ gauge boson scenario from SN1987A. It gives an order of magnitude more stringent constraint than the earlier result~\cite{Knapen:2017xzo}.
Furthermore, the stellar cooling argument on the $U(1)_{\rm B-L}$ gauge boson scenario has to be considered separately for each polarization (transverse and longitudinal), but it was overlooked previously.
We will also find the other stringent constraint from NS1987A.

The paper is organized as follows.
Sec.~\ref{sec:DPcoup} is devoted to a description of the interactions of dark gauge bosons.
We find the relevant couplings to hadrons for nucleon-nucleon bremsstrahlung, based on the symmetry argument with chiral perturbation theory.
We also discuss how dark gauge bosons interact effectively with hadrons in medium.
In Sec.~\ref{sec:DPprod}, we discuss the production of dark gauge bosons inside NS and revisit the constraints on the $U(1)_{\rm B-L}$ gauge boson scenario from SN1987A and NS1987A.
In Sec.~\ref{sec:Conclusions}, we discuss some theoretical features worthwhile to understand the points of this paper then conclude.
Appendices provide the details useful for calculations.


\section{Effective dark gauge boson couplings}
\label{sec:DPcoup}


The general Lagrangian to describe dark gauge boson interactions to the SM particles reads
\be
\mathcal{L} = -\frac{1}{4}F_{\mu\nu}F^{\mu\nu}-\frac{1}{4}F^\prime_{\mu\nu}F^{\prime\mu\nu} +\frac{\varepsilon}{2}F_{\mu\nu}F^{\prime \mu\nu} + \frac{m_{\gamma^\prime}^2}{2}A^{\prime}_\mu A^{\prime \mu}+ eA_\mu J_{\rm EM}^\mu + e^\prime A^\prime_\mu J^{\prime \mu} \, ,
\ee
where $F_{\mu\nu} = \partial_\mu A_\nu - \partial_\nu A_\mu$ and $F^\prime_{\mu\nu} = \partial_\mu A^{\prime}_\nu - \partial_\nu A^\prime_\mu$ are the field strength of the photon $A_\mu$ and the dark gauge boson $A^\prime_\mu$, respectively, and $m_{\gamma^\prime}$ is the dark gauge boson mass.
Dark gauge bosons can interact with the SM particles via the dimensionless kinetic mixing $\varepsilon$~\cite{Holdom:1985ag} with the electromagnetic (EM) current $e J^{\mu}_{\rm EM}$ and/or an explicit charge assignment in the dark $U(1)$ current $e^\prime J^{\prime \mu}$.

At scales just above the strong confinement ($\sim 1\,{\rm GeV}$) where Quantum ChromoDynamics (QCD) is still valid, the effective Lagrangian for dark gauge bosons remains the light degrees in the SM such as light quarks ($u$, $d$, and $s$), and leptons.
In this paper, though phenomenology of dark gauge bosons rely on how they interact with the SM particles, we consider a simple case that dark gauge bosons couple to the vector currents as follows
\bea
J^{\prime \mu} = \sum_{f=q,l} q_f^\prime \bar{f}\gamma^\mu f \, ,
\label{eq:DPvector}
\eea
where $q^\prime_f$ is the dark $U(1)$ gauge quantum number of the light quark fields $q = (u,d,s)$ and the lepton fields $l=(e,\mu)$.
Likewise, $q_f$ denotes the EM quantum number.

While the perturbative QCD approach breaks down at scales below the confinement scale, there is the effective theory to describe the physics in such a non-perturbative regime, known as Chiral Perturbation Theory (ChPT)~\cite{Weinberg:1978kz,Gasser:1983yg,Gasser:1984gg,Pich:1995bw}.
The prescription to come up with couplings to hadronic resonance states is based on matching a current in terms of its symmetry properties.
The dark gauge boson couplings to nucleons read
\bea
e^\prime A^\prime_\mu \sum_{N=p,n}q^\prime_N \bar{N}\gamma^\mu N
\label{eq:DPnucleon}
\eea
with $q_p^\prime = 2q^\prime_u + q^\prime_d$ and $q_n^\prime = q^\prime_u + 2q^\prime_d$.
Dark gauge bosons could also couple to charged pions as following
\be
i \left( q_p^\prime - q_n^\prime\right)e^\prime A^\prime_\mu \left[\pi^- \left(\partial^\mu \pi^+\right) - \left(  \partial^\mu \pi^-\right) \pi^+\right] \ .
\label{eq:DPpionCoup}
\ee
Furthermore, there is the leading order contact term to nucleons and charged pions similar to the Kroll–Ruderman current to photons~\cite{Kroll:1953vq}
\be
-i \left(q_p^\prime - q_n^\prime \right)e^\prime A_{\mu}^\prime \frac{f_{np}}{m_\pi}\left(\pi^+ \bar{p}\gamma^\mu \gamma^5 n - \pi^- \bar{n}\gamma^\mu \gamma^5 p\right) \, .
\label{eq:DPKRterm}
\ee
The coupling coefficient $f_{np}=\sqrt{2}f_\pi$ with the dimensionless parameter $f_\pi \sim 1$ would also appear in the pion-nucleon couplings due to the same baryon current matching.
Dark gauge boson interactions with electrons are just given by $e^\prime q_e^\prime A^\prime_\mu \bar{e}\gamma^\mu e$ as in Eq.~\eqref{eq:DPvector}.

We are interested in the dark gauge boson production inside NS, which are constituted by a hot and dense plasma.
We notice that the in-medium mixing between $A_\mu$ and $A^\prime_\mu$~\cite{Hardy:2016kme} could make an impact on effective couplings of dark gauge bosons.
In order to account for the plasma effect properly (and also a kinetic mixing via $\varepsilon$), we have to appreciate a peculiar dispersion relation of electromagnetic excitations in the medium~\cite{Braaten:1993jw,An:2013yfc,Redondo:2013lna}.
Because a dominant contribution to the dispersion relation is given by electrons, the lightest charged particles, we can approximate the effective coupling to the SM particle $f$ in medium as $\sum_f e_{\rm eff}^f A^\prime_\mu \bar{f}\gamma^\mu f $ with (for details, see Ref.~\cite{Hong:2020bxo})
\bea
e_{\rm eff}^f = e^\prime \left(q^\prime_e q_f - q_e q_f^\prime\right) + \left(\varepsilon e - e^\prime q_e^\prime\right)q_f \frac{m_{\gamma^\prime}^2}{m_{\gamma^\prime}^2 - \pi_{\rm T/L}} \, .
\label{eq:effectiveInMedium}
\eea
Here, $\pi_{\rm T/L}$ quantifies the modification of a dispersion relation for each (transverse/longitudinal) polarizations of photons in medium.
In the weak-coupling regime, the real part of  $\pi_{\rm T/L}$\footnote{There is also the imaginary part of $\pi_{\rm T/L}$, which can be interpreted as the absorption of the electromagnetic excitation with the dispersion relation of dark gauge bosons. For details, see Refs.~\cite{An:2013yfc,Chang:2016ntp}.} reads~\cite{Braaten:1993jw} 
\bea
\pi_{\rm T} =  \omega_{\rm pl}^2 \left[1+ \frac{1}{2}G\left(v_*^2 k^2/\omega^2\right)\right] \, , \qquad \quad
\pi_{\rm L}  = \omega_{\rm pl}^2 \frac{m_{\gamma^\prime}^2}{\omega^2} \frac{1-G\left(v_*^2 k^2/\omega^2\right)}{1-v_*^2 k^2/\omega^2} \, ,
\eea
where
\bea
\omega_{\rm pl}^2 = \frac{4\pi \alpha n_e}{\sqrt{m_e^2+(3\pi^2n_e)^{2/3}}}
\eea
is the plasma frequency, $v_*$ the typical electron velocity, and $\omega$ and $k$ denotes the energy and spatial momentum of an external dark gauge boson, respectively.
The function $G(x)$ is~\cite{Braaten:1993jw}
\bea
G(x) = \frac{3}{x}\left(1-\frac{2x}{3}-\frac{1-x}{2\sqrt{x}}\ln\frac{1+\sqrt{x}}{1-\sqrt{x}}\right) \, .
\eea

In this paper, we focus on the specific dark $U(1)$ gauge boson scenario that dark gauge bosons couple to the anomaly-free $\rm B-L$ current, where $\rm B$ and $\rm L$ denote the baryon and lepton number charge, respectively.
For simplicity, we ignore the kinetic mixing ($\varepsilon = 0$).\footnote{In the presence of non-vanishing kinetic mixing $\varepsilon$, we can take the effective kinetic mixing parameter including plasma mixing as $\varepsilon_{\rm eff}=\varepsilon - e^\prime q^\prime_e/e$ in Eq.~\eqref{eq:effectiveInMedium}.}
We then find the effective couplings in the form of $\sum_{f=e,p,n} e^\prime \tilde{q}_{f}^{\prime}  A^\prime_\mu \bar{f}\gamma^\mu f $ with 
\bea
\tilde{q}_{e}^{\prime} & = & - \frac{m_{\gamma^\prime}^2}{m_{\gamma^\prime}^2 - \pi_{\rm T/L}} \, , \label{eq:EffCoupElectron}\\
\tilde{q}_{p}^{\prime} & = &  \frac{m_{\gamma^\prime}^2}{m_{\gamma^\prime}^2 - \pi_{\rm T/L}} \, ,\\
\tilde{q}_{n}^{\prime} & = & 1 \, .
\eea
Moreover, we can directly put these effective couplings to nucleons into the couplings in \Eqs{eq:DPpionCoup}{eq:DPKRterm} then
\bea
q_p^\prime - q_n^\prime \rightarrow \tilde{q}_{p}^{\prime}-\tilde{q}_{n}^{\prime} = \frac{\pi_{\rm T/L}}{m_{\gamma^\prime}^2 - \pi_{\rm T/L}} \, .
\label{eq:EffCoupqp-qn}
\eea
When plasma screening is large enough, i.e. $\pi_{\rm T/L} \gg m_{\gamma^\prime}^2$, the difference of the effective couplings to nucleons in medium becomes of ${\cal O}(1)$.
As we will discuss later, this  isospin breaking combination leads to enhancement of the emission rate of $U(1)_{\rm B-L}$ gauge bosons in the context of an expansion of the order parameter.


\section{Dark gauge boson production via Nucleon-Nucleon bremsstrahlung}
\label{sec:DPprod}


Let us discuss the emission rate of dark gauge bosons via nucleon bremsstrahlung.
Since nucleons are semi-relativistic even under such extreme circumstances in the core of NS, a velocity of nucleons is still a good expansion parameter for nucleon scatterings~\cite{Rrapaj:2015wgs}.
For dark gauge boson, we neglect its mass in calculations.
This assumption is feasible only when $m_{\gamma^\prime}$ is much smaller than any other thermodynamic variables such as temperature.
While the mass effect becomes important for a larger value $m_{\gamma^\prime} \geq \mathcal{O}(T)$, at the same time, the exponential phase space suppression $e^{-m_{\gamma^\prime}/T}$ appears in the dark gauge boson emissitivty. The resulting constraint vanishes prominently.

\begin{figure}[t!]
\centering
\includegraphics[width=0.95\textwidth]{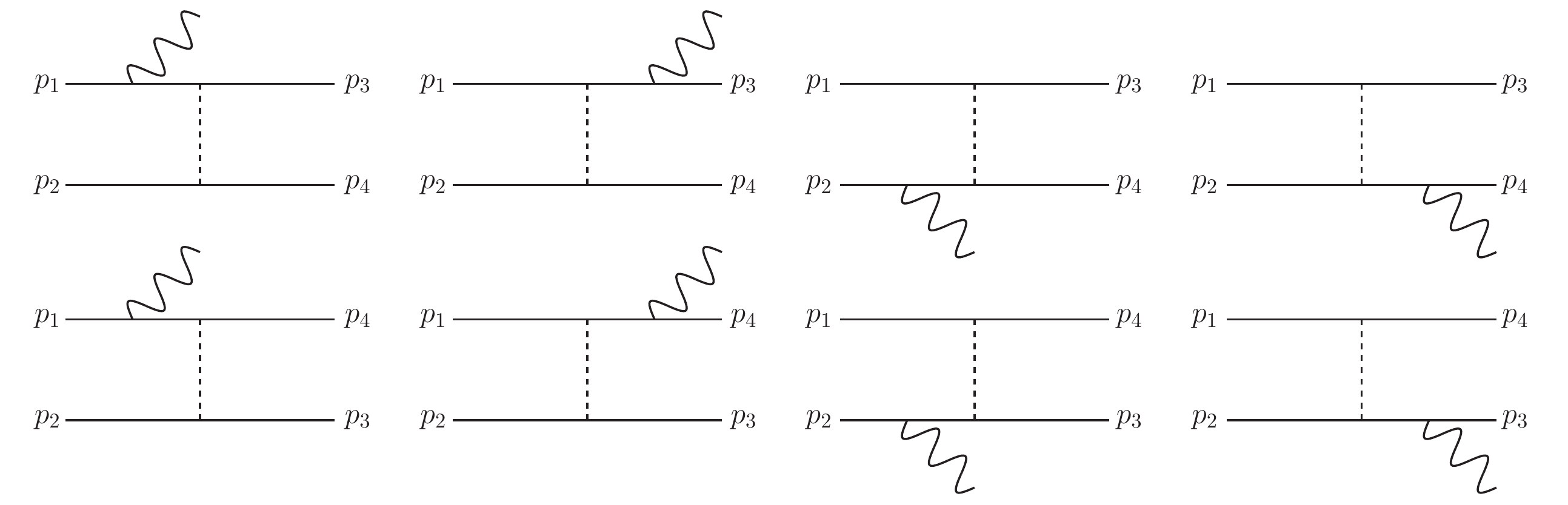}
\caption{\em  Diagrams of nucleon-nucleon bremsstrahlung with the exchange of one pion (dashed line), that the dark gauge boson (wavy line) attaches to external nucleon legs (solid line) with the corresponding momentum assignment.}
\label{fig:NNbrem}
\end{figure}

For nucleon bremsstrahlung, we employ the approximation of one-pion-exchange that nucleons interact themselves via an one-pion mediator.
According to ChPT, the leading order nucleon-pion couplings read
\be
\frac{f_{NN}}{m_\pi}\partial_\mu \pi^0 \bar{N}\gamma^\mu \gamma^5 N + \frac{f_{N^\prime N}}{m_\pi} \partial_\mu \pi^\pm \bar{N}^\prime \gamma^\mu \gamma^5 N \, ,
\ee
where $N =(p,n)$ denotes the nucleon field, and $N^\prime=(n, p)$ is the different nucleon with respect to $N$.
Due to $SU(2)$ isospin structure, there is the relation between the coupling coefficients $f_{pp} = -f_{nn} =f_{np}/\sqrt{2} = f_{pn}/\sqrt{2} = f_\pi$.

In the case of bremsstrahlung from the identical nucleon scatterings, e.g. $pp\rightarrow pp\gamma^\prime$ or $nn\rightarrow nn \gamma^\prime$, dark gauge bosons are produced through its couplings to nucleons as in Fig.~\ref{fig:NNbrem}.
We have totally 8 diagrams for these processes.
The dashed and wavy lines denote the pion mediator and the dark gauge boson, respectively.
Nucleons are represented by the solid lines, and we define the four momentum of the two incident nucleons as $p_{1}$ and $p_2$, and that of the two scattered nucleons as $p_{3}$ and $p_4$.
Since nucleons are indistinguishable, the scattering amplitudes are anti-symmetric under exchange of $p_3$ and $p_4$.

Scattering of neutrons by protons, and vice versa, can also contribute to the production of dark gauge bosons, and this corresponds to $n$-$p$ bremsstrahlung $np\rightarrow np \gamma^\prime$.
This process may turn out to be the leading order emission of dark gauge bosons if the (effective) dark charge assignment to neutrons and protons is asymmetric~\cite{Rrapaj:2015wgs}.
There are also the 8 diagrams shown in Fig.~\ref{fig:NNbrem} in common with identical nucleon bremsstrahlung cases.
Without loss of generality, we set the nucleon with $p_1$ equate to that with $p_3$, and the same for $p_2$ and $p_4$.
Then the four upper diagrams in Fig.~\ref{fig:NNbrem} correspond to the $t$-channel processes, connecting the respective nucleons to each other through a neutral pion mediator.
The other four lower diagrams in Fig.~\ref{fig:NNbrem} are the $u$-channel processes, which exchange the external nucleon leg of $p_3$ with of $p_4$ in comparison with the $t$-channel processes, and mediated by a charged pion.
 
\begin{figure}[t!]
\centering
\includegraphics[width=0.8\textwidth]{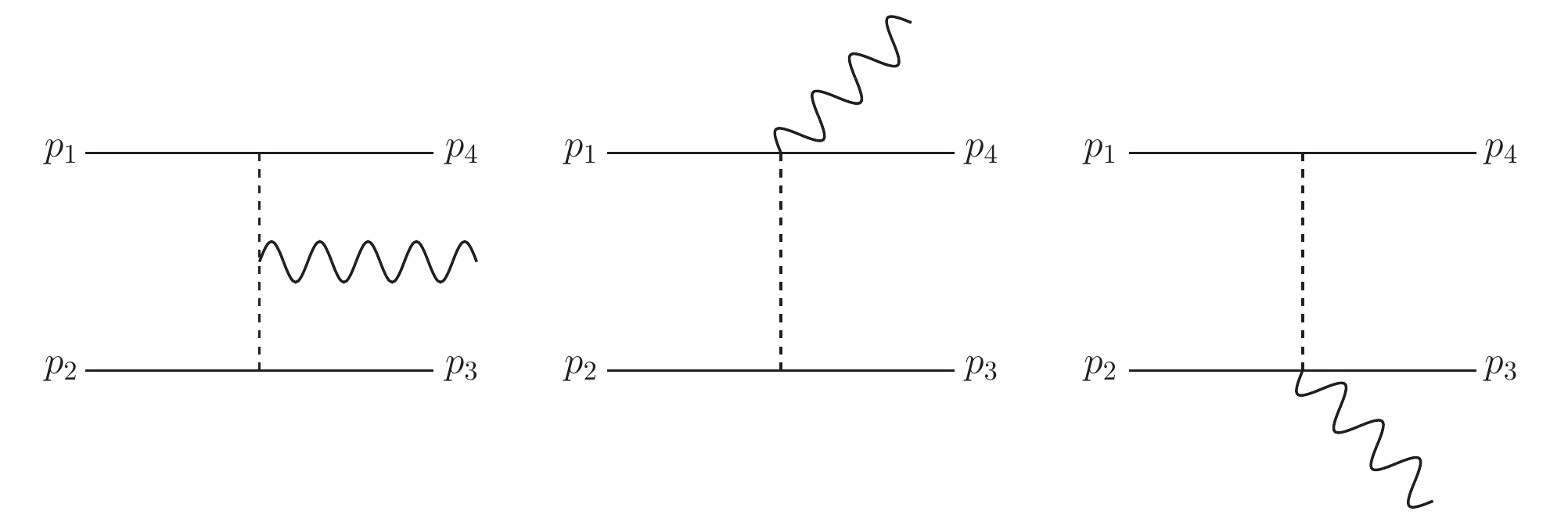}
\caption{\em  Additoinal diagrams for $n$-$p$ bremsstrahlung. Notation as in Fig.~\ref{fig:NNbrem}.}
\label{fig:NNbremIC}
\end{figure}

Furthermore, the dark gauge boson couplings to charged pions in Eq.~\eqref{eq:DPpionCoup} and \eqref{eq:DPKRterm} induce the three additional diagrams for $n$-$p$ bremsstrahlung as presented in Fig.~\ref{fig:NNbremIC}.
The first diagram in Fig.~\ref{fig:NNbremIC} indicates the `internal' bremsstrahlung of a charged pion mediator through the coupling given in Eq.~\eqref{eq:DPpionCoup}.
The dark gauge boson leg for the next two diagrams attaches to a vertex of nucleons and a charged pion due to the contact interaction in Eq.~\eqref{eq:DPKRterm}.
We dub these processes as the `contact' bremsstrahlung.
These diagrams are non-vanishing when protons and neutrons carry distinguishable effective charges on the dark $U(1)$ gauge (i.e., $\tilde{q}^\prime_n \neq \tilde{q}^\prime_p$), and may contribute to the leading order emission of dark gauge bosons.

\subsection{Matrix elements}
\label{sec:MatrixElements}

There are plenty of terms in the matrix elements (and consequently in the square of them), but one can simplify them by taking expansion of function in an order parameter.
We provide the matrix elements at each orders in terms of multipole expansion, which is the proper treatment in a non-relativistic nucleon gas limit.
Namely, we consider a velocity of nucleons as the order parameter for expansion; more specifically, a momentum transfer of nucleons.
We denote $\mathbf{k} \equiv \vec{p}_2-\vec{p}_4 = \vec{p}_3-\vec{p}_1 + \vec{p}_{\gamma^\prime} \simeq \vec{p}_3 - \vec{p}_1$ and $\mathbf{l} \equiv \vec{p}_2-\vec{p}_3 = \vec{p}_4-\vec{p}_1 + \vec{p}_{\gamma^\prime} \simeq \vec{p}_4 - \vec{p}_1$, where $\vec{p}_i$ is the spatial momentum of $p_i$.

The matrix element for the four upper diagrams in Fig.~\ref{fig:NNbrem} corresponding to the $t$-channel process reads
\be
\begin{split}
i\mathcal{M}_{t} & =  ie^\prime  \left(\frac{f_{13}}{m_\pi}\right)\left(\frac{f_{24}}{m_\pi}\right) \frac{1}{\left(p_2-p_4\right)^2-m_\pi^2} \, \left(-2m_N\right) \bar{u}(p_4) \gamma^5 u(p_2)  \\
& \qquad \times \bar{u}(p_3)\left[\,  \tilde{q}_3^\prime \slashed{\epsilon}\frac{ \left(\slashed{p}_3+\slashed{p}_{\gamma^\prime} + m_N\right)}{2 p_3\cdot p_{\gamma^\prime}+m_{\gamma^\prime}^2}  \left(\slashed{p}_3-\slashed{p}_{1}+\slashed{p}_{\gamma^\prime}\right)\gamma^5  \right.  \\
& \qquad \qquad \quad \left. + \tilde{q}_1^\prime \left(\slashed{p}_3-\slashed{p}_{1}+\slashed{p}_{\gamma^\prime}\right)\gamma^5 \frac{ \slashed{p}_1-\slashed{p}_{\gamma^\prime} + m_N}{-2 p_1\cdot p_{\gamma^\prime}+m_{\gamma^\prime}^2}   \slashed{\epsilon} \right]u(p_1)  \\
& + \Big\{p_1 \leftrightarrow p_2 , p_3 \leftrightarrow p_4, \tilde{q}_1^\prime \leftrightarrow \tilde{q}_2^\prime , \tilde{q}_3^\prime \leftrightarrow \tilde{q}_4^\prime \Big\} \, ,
\end{split}
\label{eq:tChannel_N-DP}
\ee
where $u(p_i)$ denotes the spinor of a nucleon with its four-momentum $p_i$ and the effective coupling in medium $\tilde{q}_i^\prime$,  and $\epsilon_\mu$ is the polarization vector of dark gauge bosons, and 
the first term in Eq.~\eqref{eq:tChannel_N-DP} arises from the two left diagrams of upper ones in Fig.~\ref{fig:NNbrem}.
The matrix element for the other two diagrams can be written as the first term with the interchange rule given in the last term in Eq.~\eqref{eq:tChannel_N-DP}.
Due to  $\tilde{q}_1^\prime = \tilde{q}_3^\prime$ and $\tilde{q}_2^\prime = \tilde{q}_4^\prime$, we can simplify $i\mathcal{M}_{t}$ as
\be
i\mathcal{M}_{t} \simeq i \mathcal{M}_{t(J,1)} +  i \mathcal{M}_{t(J,2)} + i \mathcal{M}_{t(\sigma,2)}
\ee
with
\begin{eqnarray}
i\mathcal{M}_{t(J,1)}  & = & \,  ie^\prime \left(\tilde{q}_1^\prime - \tilde{q}_2^\prime\right)  \frac{f_{13}f_{24}}{m_\pi^2}  \frac{1}{\mathbf{k}^2+m_\pi^2} \frac{4m_N}{\omega}\,  \bar{u}(p_4) \gamma^5 u(p_2)\, \bar{u}(p_3) \gamma^5 u(p_1)  \, \left[ \epsilon_\mu J^\mu_{(1)} \right] \, , \label{eq:MtJ1}\\
i\mathcal{M}_{t(J,2)}  & = & \, ie^\prime \left(\tilde{q}_1^\prime + \tilde{q}_2^\prime\right)   \frac{f_{13}f_{24}}{m_\pi^2}  \frac{1}{\mathbf{k}^2+m_\pi^2}   \frac{2m_N}{\omega} \, \bar{u}(p_4) \gamma^5 u(p_2)\, \bar{u}(p_3) \gamma^5 u(p_1)  \, \left[- \epsilon_\mu J^\mu_{(2)} \right] \, , \ \ \label{eq:MtJ2}\\
i\mathcal{M}_{t(\sigma,2)}  & = & \, ie^\prime \tilde{q}_1^\prime  \frac{f_{13}f_{24}}{m_\pi^2}  \frac{1}{\mathbf{k}^2+m_\pi^2} \frac{2m_N}{\omega} \, \bar{u}(p_4) \gamma^5 u(p_2) \bar{u}(p_3)\sigma_{\mu\nu}\gamma^5 u(p_1) \cdot i p_{\gamma^\prime}^{\mu} \epsilon^\nu \frac{\mathbf{k}\cdot\hat{p}_{\gamma^\prime}}{m_N}\nn  \\
&& + \Big\{p_1 \leftrightarrow p_2 , p_3 \leftrightarrow p_4, \tilde{q}_1^\prime \leftrightarrow \tilde{q}_2^\prime , \mathbf{k} \rightarrow -\mathbf{k}\Big\} \, , \label{eq:MsJ1}
\end{eqnarray}
where $\sigma^{\mu\nu} = i \left[\gamma^\mu, \gamma^\nu\right]/2$ and $\hat{p}_{\gamma^\prime} = \vec{p}_{\gamma^\prime}/|\vec{p}_{\gamma^\prime}$.
Here,  $J_{(1)}^\mu$ and $J_{(2)}^\mu$ correspond to the dipole and quadrupole current~\cite{Rrapaj:2015wgs}, respectively, and they are written as
\be
J^\mu_{(1)}  =  \left\{\mathbf{k}\cdot \hat{p}_{\gamma^\prime}, \mathbf{k}\right\} \, , \qquad   J^\mu_{(2)}   =  \frac{1}{m_N} \left\{2 \left(\mathbf{k}\cdot \hat{p}_{\gamma^\prime} \right) \left(\mathbf{l}\cdot \hat{p}_{\gamma^\prime}\right), \mathbf{k}\left(\mathbf{l}\cdot \hat{p}_{\gamma^\prime}\right) +\mathbf{l}\left(\mathbf{k}\cdot \hat{p}_{\gamma^\prime}\right) \right\} \, .
\ee
Due to $\left[ p_{\gamma^\prime}\right]_\mu J_{(i)}^\mu = 0$, one can easily check that the above matrix elements  follow the Ward-Takahashi identity, which implies the conservation of the vector currents.

The subscript of matrix elements is in the form of $a(\alpha,i)$, and it indicates as follows: the `i'-th order matrix element for `a'-channel process in `$\alpha$'-type.
One can easily check the order expansion in powers of the ratio of a nucleon momentum transfer to the nucleon mass that says $|\mathcal{M}_{a(\alpha,2)}|/|\mathcal{M}_{a(\alpha,1)}| \sim \mathcal{O}(|\mathbf{k}|/m_N)$ with regard to $t$-channel processes. For $u$-channel processes, it is $\mathcal{O}(|\mathbf{l}|/m_N)$.
 Note that all the leading order matrix elements $\mathcal{M}_{a(\alpha,1)}$ is proportional to the difference between the effective couplings to the two incident nucleons, hence the only $n$-$p$ bremsstrahlung scatterings can radiate dark gauge bosons with a leading order emission rate.
In $J$ and $\sigma$-type, the polarization vector of a dark gauge boson couples to the $J_{(i)}^\mu$ current and the dipole moment of nucleons, respectively.
As we will discuss, the $J$-type contains the 1-less power of $\omega$ than the $\sigma$-type at the same order.
This implies that at the soft radiation approximation~\cite{Low:1958sn,Nyman:1968jro,Heller:1968cry,Rrapaj:2015wgs} (i.e., $\omega \ll E_{\rm cm}$ where $E_{\rm cm}$ is the kinetic energy in the non-relativistic center of mass frame of nucleons) one can neglect $\sigma$-type terms as adopted in the literature for discussion on the dark photon scenario~\cite{Rrapaj:2015wgs,Chang:2016ntp}.

Since the $u$-channel processes of the four lower diagrams in Fig.~\ref{fig:NNbrem} are anti-symmetric to the $t$-channel under $p_3\leftrightarrow p_4$ with $\tilde{q}^\prime_3 \leftrightarrow \tilde{q}^\prime_4$, one can straightforwardly derive its matrix element from the result of $i\mathcal{M}_t$
\be
\begin{split}
i\mathcal{M}_{u} & = - ie^\prime  \left(\frac{f_{14}}{m_\pi}\right)\left(\frac{f_{23}}{m_\pi}\right) \frac{1}{\left(p_2-p_3\right)^2-m_\pi^2} \, \left(-2m_N\right) \bar{u}(p_3) \gamma^5 u(p_2)  \\
& \qquad \times \bar{u}(p_4)\left[\,  \tilde{q}_4^\prime \slashed{\epsilon}\frac{ \left(\slashed{p}_4+\slashed{p}_{\gamma^\prime} + m_N\right)}{2 p_4\cdot p_{\gamma^\prime}+m_{\gamma^\prime}^2}  \left(\slashed{p}_4-\slashed{p}_{1}+\slashed{p}_{\gamma^\prime}\right)\gamma^5  \right.  \\
& \qquad \qquad \quad \left. + \tilde{q}_1^\prime \left(\slashed{p}_4-\slashed{p}_{1}+\slashed{p}_{\gamma^\prime}\right)\gamma^5 \frac{ \slashed{p}_1-\slashed{p}_{\gamma^\prime} + m_N}{-2 p_1\cdot p_{\gamma^\prime}+m_{\gamma^\prime}^2}   \slashed{\epsilon} \right]u(p_1)  \\
& + \Big\{p_1 \leftrightarrow p_2 , p_3 \leftrightarrow p_4, \tilde{q}_1^\prime \leftrightarrow \tilde{q}_2^\prime , \tilde{q}_3^\prime \leftrightarrow \tilde{q}_4^\prime \Big\} \, .
\end{split}
\label{eq:uChannel_N-DP}
\ee
Here, the overall $(-1)$ factor is introduced due to the anti-permutation relation of nucleons.
With the same simplification for $i\mathcal{M}_t$ in Eqs.~\eqref{eq:MtJ1}-\eqref{eq:MsJ1}, we find
\be
i\mathcal{M}_{u} \simeq i \mathcal{M}_{u(J,1)} +  i \mathcal{M}_{u(J,2)} + i \mathcal{M}_{u(\sigma,1)} + i \mathcal{M}_{u(\sigma,2)}
\ee
with
\begin{eqnarray}
i\mathcal{M}_{u(J,1)}  & = & \, ie^\prime \left(\tilde{q}_1^\prime - \tilde{q}_2^\prime\right)  \frac{f_{14}f_{23}}{m_\pi^2}  \frac{1}{\mathbf{l}^2+m_\pi^2} \frac{4m_N}{\omega}\,  \bar{u}(p_3) \gamma^5 u(p_2)\, \bar{u}(p_4) \gamma^5 u(p_1)  \, \left[- \epsilon_\mu J^\mu_{(1)} \right] \, , \ \ \\
i\mathcal{M}_{u(J,2)}  & = & \, ie^\prime \left(\tilde{q}_1^\prime + \tilde{q}_2^\prime\right)   \frac{f_{14}f_{23}}{m_\pi^2}  \frac{1}{\mathbf{l}^2+m_\pi^2}   \frac{2m_N}{\omega} \, \bar{u}(p_3) \gamma^5 u(p_2)\, \bar{u}(p_4) \gamma^5 u(p_1)  \, \left[ \epsilon_\mu J^\mu_{(2)} \right] \, ,  \\
i\mathcal{M}_{u(\sigma,1)}  & = & \, ie^\prime  \left(\tilde{q}_1^\prime - \tilde{q}_2^\prime\right)   \frac{f_{14}f_{23}}{m_\pi^2}  \frac{1}{\mathbf{l}^2+m_\pi^2} \frac{2m_N}{\omega} \, \bar{u}(p_3) \gamma^5 u(p_2) \bar{u}(p_4)\sigma_{\mu\nu}\gamma^5 u(p_1) \cdot i p_{\gamma^\prime}^{\mu} \epsilon^\nu \nn  \\
&& + \Big\{p_1 \leftrightarrow p_2 , p_3 \leftrightarrow p_4, \tilde{q}_1^\prime \leftrightarrow \tilde{q}_2^\prime , \mathbf{l} \rightarrow -\mathbf{l}\Big\} \, ,\\
i\mathcal{M}_{u(\sigma,2)}  & = & \, ie^\prime \left(\tilde{q}_1^\prime + \tilde{q}_2^\prime\right)    \frac{f_{14}f_{23}}{m_\pi^2}  \frac{1}{\mathbf{l}^2+m_\pi^2} \frac{m_N}{\omega} \, \bar{u}(p_3) \gamma^5 u(p_2) \bar{u}(p_4)\sigma_{\mu\nu}\gamma^5 u(p_1) \cdot \left(-i\right) p_{\gamma^\prime}^{\mu} \epsilon^\nu \frac{\mathbf{l}\cdot\hat{p}_{\gamma^\prime}}{m_N}\nn  \\
&& + \Big\{p_1 \leftrightarrow p_2 , p_3 \leftrightarrow p_4, \tilde{q}_1^\prime \leftrightarrow \tilde{q}_2^\prime , \mathbf{l} \rightarrow -\mathbf{l}\Big\} \, .
\end{eqnarray}
We notice that these matrix elements of $i\mathcal{M}_u$ satisfy the Ward-Takahashi identity at each order as in $i\mathcal{M}_t$.

The matrix element for the `internal' bremsstrahlung of the left diagram in Fig.~\ref{fig:NNbremIC} reads
\be
\begin{split}
i\mathcal{M}_{u({\rm int},1)}  \simeq & \,  ie^\prime \left(\tilde{q}_1^\prime - \tilde{q}_2^\prime\right)  \left(\frac{f_{14}}{m_\pi}\right)\left(\frac{f_{23}}{m_\pi}\right)  8 m_N^3 \frac{\mathbf{l} \cdot \vec{\epsilon}}{m_N}    \\
& \times \frac{1}{\mathbf{l}^2+m_\pi^2}\frac{1}{\mathbf{l}^2-2\mathbf{l}\cdot \vec{p}_{\gamma^\prime}+m_\pi^2}\, \bar{u}(p_4) \gamma^5 u(p_1) \bar{u}(p_3) \gamma^5 u(p_2)  \, .
\end{split}
\ee
Likewise, we also have the matrix element for the `contact' bremsstrahlung of the next two diagrams in Fig.~\ref{fig:NNbremIC}
\be
\begin{split}
i\mathcal{M}_{u({\rm con},1)}  \simeq & \,  -ie^\prime \left(\tilde{q}_1^\prime - \tilde{q}_2^\prime \right)  \left(\frac{f_{14}}{m_\pi}\right)\left(\frac{f_{23}}{m_\pi}\right)   \,  2m_N\\
& \times \epsilon_\mu  \left[ \frac{1}{\mathbf{l}^2+m_\pi^2} \bar{u}(p_4) \gamma^\mu \gamma^5 u(p_1) \bar{u}(p_3) \gamma^5 u(p_2)  \right. \\
& \qquad \quad \left. - \frac{1}{\mathbf{l}^2 - 2 \mathbf{l}\cdot \vec{p}_{\gamma^\prime} + m_\pi^2} \bar{u}(p_4) \gamma^5 u(p_1) \bar{u}(p_3)\gamma^\mu  \gamma^5 u(p_2) \right] \, .
\end{split}
\ee
Here, $\mathbf{l}\cdot \vec{p}_{\gamma^\prime}$ factor remains in the denominator to account for fulfilment of the Ward-Takahashi identity.
These terms will be neglect as a good approximation when we perform calculations.

\subsection{Emission rates}

The volume emissivity (i.e. emission rate per volume) of dark gauge bosons via nucleon-nucleon bremsstrahlung is written as
\be
\begin{split}
Q_{\rm Brem}  = & S \left[ \prod_{i=1}^4 \int \frac{d \vec{p}_i}{2E_i(2\pi)^3} \right] \int \frac{d \vec{p}_{\gamma^\prime}}{2\omega(2\pi)^3} \cdot \omega \left| \mathcal{M}_{\rm emi}\right|^2 e^{-\tau}f_1 f_2 \left(1-f_3\right)\left(1-f_4\right)  \\
& \qquad \qquad \qquad \qquad \qquad \times \left(2\pi\right)^4\delta^{(4)} \left(p_1 + p_2 - p_3 - p_4 - p_{\gamma^\prime}\right) \, ,
\label{eq:EmissivityEq}
\end{split}
\ee
where $S$ denotes the symmetry factor (if nucleons are identical, $1/4$; otherwise $1$),  $f_i = \left(\exp\left[\left(E_i - \mu_{F_i}\right)/T\right] + 1\right)^{-1}$ the Fermi-Dirac distribution function with a chemical potential $\mu_{F_i}$, and $\mathcal{M}_{\rm emi}$ the total scattering amplitude for the dark gauge boson emission.
All the matrix elements at each order with the one-pion-exchange approximation are derived in the previous subsection, and the squares of them  (including cross terms) with the polarization sum are provided in App.~\ref{app:SqMatrix}.
We also includes the optical depth ($\equiv \tau$) of a dark gauge boson in the Eq.~\eqref{eq:EmissivityEq} as well.
Due to $\tau \propto e^{\prime 2}$, dark gauge bosons with a very small $e^\prime$ freely stream out of NS, whereas they may be trapped (or absorbed) significantly inside NS for a large coupling regime then net luminosity of dark gauge bosons could be suppressed.

First, let us discuss the $U(1)_{\rm B-L}$ gauge boson production from a proto-neutron star during supernovae explosion.
Since the nuclear matter in the core of supernovae is partially degenerate $\mu_{F_i}/T \approx 1$,  we take the plausible assumption that the nucleons in the core of supernovae are non-degenerate as well as non-relativistic, which is the so-called `classical' limit.
The phase space integral in the classical limit is summarized in App.~\ref{app:PSclassical}.
We point out $\left<\mathbf{k}^2/2m_N\right> \sim \left<\mathbf{l}^2/2m_N\right> \sim \left<\omega\right> \sim T$ where the bracket indicates a thermal average.
Thus not only the $J$-type matrix elements, which are relevant for the soft radiation approximation~\cite{Low:1958sn,Nyman:1968jro,Heller:1968cry}, but also the others, such as the $\sigma$-type, give comparable contributions to the total scattering amplitude at each order.
To calculate the emission rate, we employ a simple analytic profile of the proto-neutron star at 1 second after core bounce, which is dubbed as a `fiducial' model in Ref.~\cite{Chang:2016ntp}:
\be
\begin{split}
\rho\left(r\right) & = \rho_c \times\left\{
\begin{tabular}{cc}
$1+\kappa_\rho \left(1-r/R_c\right)$ & $r < R_c$\\
$\left(r/R_c\right)^{-\nu}$ & $r \geq R_c$
\end{tabular}
\right.  \, , \\
T\left(r\right) & =   T_c \times\left\{
\begin{tabular}{cc}
$1+\kappa_T \left(1-r/R_c\right)$ & $r < R_c$\\
$\left(r/R_c\right)^{-\nu/3}$ & $r \geq R_c$
\end{tabular}
\right. \, 
\end{split}
\label{eq:fiducialModel}
\ee
with the core radius $R_c =10\,{\rm km}$, the core density $\rho_c = 3\times 10^{14}\,{\rm g}/{\rm cm}^{3}$, the core temperature $T_c =  30\,{\rm MeV}$, $\kappa_\rho = 0.2$, $\kappa_T = -0.5$, $\nu = 5$, and a uniform proton fraction $Y_p= 0.3$.
$U(1)_{\rm B-L}$ gauge bosons are produced dominantly from the core. Outside the core, the density and temperature profile are suppressed steeply by a power-law ansatz with the negative exponent~\cite{Raffelt:1996wa}.

Because of the in-medium mixing~\cite{Hardy:2016kme}, protons carry a different effective gauge charge of $U(1)_{\rm B-L}$ compared to that of neutrons. 
  As described in Sec.~\ref{sec:MatrixElements}, the leading order matrix elements of $n$-$p$ bremsstrahlung process, which correspond to $\mathcal{M}_{a(\alpha,1)}$ for any `a'-channel and `$\alpha$'-type, is controlled by the difference between the effective couplings to neutrons and protons as given in Eq.~\eqref{eq:EffCoupqp-qn}.
Thus, $U(1)_{\rm B-L}$ gauge bosons are mainly produced in $n$-$p$ bremsstrahlung at the leading order even though the nucleon fields in the vacuum have the degenerate charge under the $U(1)_{\rm B-L}$ gauge symmetry.
Moreover, such an effective coupling depends on the momentum of emitted $U(1)_{\rm B-L}$ gauge bosons~\cite{Braaten:1993jw}.
Inside the core of the proto-neutron star, $\omega_{\rm pl}\sim T$ so that a resonance condition $\pi_{\rm T/L} = m_{\gamma^\prime}^2$ can be achieved to enhance the emission rate~\cite{An:2013yfc,Chang:2016ntp}.
However, it is verified that the $\omega$ integral of the differential power around a resonance gives a sub-dominant contribution to the total luminosity.
 For the transverse polarizations, a resonance can occur only when the mass is rather heavy as $\omega_{\rm pl} \leq m_{\gamma^\prime} \leq \sqrt{3/2}\, \omega_{\rm pl} \simeq 1.225\omega_{\rm pl}$.
The imaginary part of $\pi_{\rm T}$ at the resonance is an $\mathcal{O}(1)$ larger than $m_{\gamma^\prime}^2$, a broad resonance width  is imposed to undermine a resonance effect (for more details, see Ref.~\cite{Chang:2016ntp}).
In a small mass region $m_{\gamma^\prime}<\omega_{\rm pl}$, longitudinal polarizations of the photon excitations can always have a dispersion relation to reach a resonance.
The energy spectrum of longitudinal $U(1)_{\rm B-L}$ gauge bosons typically contains a narrow peak at the resonance due to the small value of ${\rm Im}\,\pi_{\rm L}$. But its contribution to the total luminosity of the longitudinal modes is suppressed due to a rather large resonance energy in comparison with the temperature.
As a result, the prescription to ignore the effective coupling to protons (i.e., $| \tilde{q}_{p}^{\prime} - \tilde{q}_{n}^{\prime} | \simeq 1$) in the small mass regime for $m_{\gamma^\prime} <\omega_{\rm pl}$ is in a good agreement with the actual results.
For $U(1)_{\rm B-L}$ gauge bosom masses larger than $\omega_{\rm pl}$, the plasma screening leads to $| \tilde{q}_{p}^{\prime} - \tilde{q}_{n}^{\prime} | \propto m_{\gamma^\prime}^{-2}$.

On the other hands, the produced $U(1)_{\rm B-L}$ gauge bosons can be reabsorbed into the medium of the proto-neutron star when they interact strongly enough leading to a large value of the optical length $\tau$.
For small masses, the major absorption arises from inverse bremsstrahlung.
The absorption rate via inverse bremsstrahlung can be written as similar to the emission rate in Eq.~\eqref{eq:EmissivityEq} except the integral over $d\vec{p}_{\gamma^\prime}$ and the additional energy $\omega$ 
\be
\begin{split}
\Gamma_{\rm invBrem}  = & \frac{1}{2\omega} S \left[ \prod_{i=1}^4 \int \frac{d \vec{p}_i}{2E_i(2\pi)^3} \right]  \left| \mathcal{M}_{\rm abs}\right|^2 f_1 f_2 \left(1-f_3\right)\left(1-f_4\right)  \\
& \qquad \qquad \qquad \qquad \qquad \times \left(2\pi\right)^4\delta^{(4)} \left(p_1 + p_2 + p_{\gamma^\prime} - p_3 - p_4\right) \, .
\label{eq:AbsorptionEq}
\end{split}
\ee
Here, the total scattering amplitude for absorption $\mathcal{M}_{\rm abs}$ consists of the matrix elements given in Sec.~\ref{sec:MatrixElements}, but with the different sign for the $J$-type ones because the $U(1)_{\rm B-L}$ gauge boson is now an incident particle (i.e., the different sign of its four-momentum in the energy-momentum conservation compared to emission cases).
The absorption of $U(1)_{\rm B-L}$ gauge bosons with a mass as $m_{\gamma^\prime} > 10\,{\rm MeV}$ is dominated by the decay to electron-positron pairs $\gamma^\prime \rightarrow e^- e^+$~\cite{Chang:2016ntp}.
The effective coupling to electrons in medium as given by Eq.~\eqref{eq:EffCoupElectron} is equivalent to that for dark photons by replacing $\varepsilon e \rightarrow e^\prime$. Thus, one can use the result of the dark photon scenario~\cite{Chang:2016ntp} in order to account for the decay rate.
There are certain technical obstacles to the calculation of the optical length $\tau$, because 
we have to take into account the averaged integral of the absorption rate along all the possible trajectories at each production point.
To deal with these difficulties, one can simplify the situations for the absorption process, such as a spherical symmetric profile.
We follow the well formulated procedure in Ref.~\cite{Chang:2016ntp}.
This methodology introduces the specific radius, inside of which the absorbed novel particle's energy can be reprocessed into thermal neutrino energy.
This radius is dubbed as `far radius'. It is literally taken far from the neutrino sphere ($R_\nu$) to ensure a definite energy absorption for a large $U(1)_{\rm B-L}$ gauge  coupling regime.
We take the conservative value of the far radius as $1000\,{\rm km}$ corresponding to the bounce shock radius.

\begin{figure}[t!]
\centering
\includegraphics[width=0.75\textwidth]{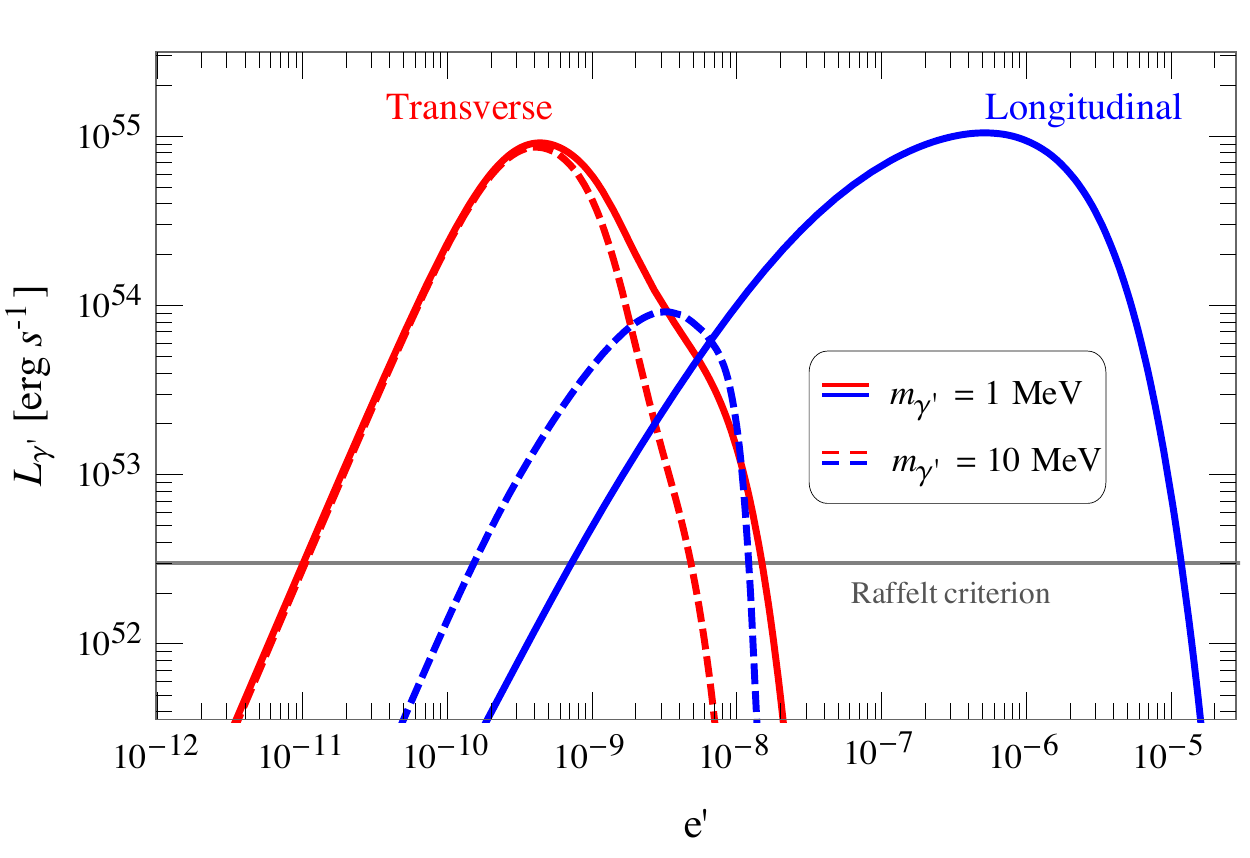}
\caption{\em Luminosities of  transverse (red) and longitudinal (blue) $U(1)_{\rm B-L}$ gauge bosons as functions of the gauge coupling $e^\prime$ for $m_{\gamma^\prime} = 1\,{\rm MeV}$ (solid) and $m_{\gamma^\prime}=10\,{\rm MeV}$ (dashed). The gray line denotes the ``Raffelt criterion'' to constrain the rate of novel particle emission during the first  few seconds of SN1987A.}
\label{fig:Sn1987Acooling}
\end{figure}

Fig.~\ref{fig:Sn1987Acooling} shows  the luminosity of transverse (red) and longitudinal (blue) $U(1)_{\rm B-L}$ gauge bosons as functions of $e^\prime$ for $m_{\gamma^\prime} = 1\,{\rm MeV}$ (solid) and $m_{\gamma^\prime}=10\,{\rm MeV}$ (dashed).
For a small coupling regime, $U(1)_{\rm B-L}$ gauge bosons freely stream out and therefore the luminosity is proportional to $e^{\prime 2}$.
The production of transverse $U(1)_{\rm B-L}$ gauge bosons in the free-streaming regime is nearly independent of $m_{\gamma^\prime}$ in the range of interest ($m_{\gamma'} <T\sim \omega_{\rm pl} \sim 10\,{\rm MeV}$) due to the plasma suppression on the effective coupling to protons, whereas the emission rate of longitudinal components is proportional to the $U(1)_{\rm B-L}$ gauge boson mass, which originates in the current conservation law.
At masses above the temperature, such a rate is exponentially suppressed by the Boltzmann factor (as well as weakened plasma screening).
Once the coupling strength gets stronger as the mean free path comparable to the geometric size of the proto-neutron star, the absorption processes work efficiently so the net energy transport of $U(1)_{\rm B-L}$ gauge bosons becomes more and more decreased.
Further reduction of the luminosity for a large mass above the $2m_e$ threshold is accounted for by the decay channel into the electron-positron pair.



We now turn to the creation of $U(1)_{\rm B-L}$ gauge bosons inside NS as a remnant of supernovae explosion.
In particular, young NS with ages under $10^4$-$10^5$ yr are of interest in terms of the stellar cooling argument because its interior is still hot for the volume emission of neutrinos to dominate the cooling.
The nuclear matter within NS is highly degenerate as $\mu_{F_i} = p_{F_i}^2/2m_N \gg T$ with $p_{F_i}$ the Fermi momentum of the nucleon $i$.
Under this circumstance of the so-called `degenerate' limit, the plasma frequency $\omega_{\rm pl}$  $(\sim 10\,{\rm MeV})$ is typically much larger than the temperature.
Hence, we can neglect the effective coupling to protons as in the supernovae case, and $n$-$p$ bremsstrahlung turns out to be the leading order process.
We summarize how to treat the phase space integral in the degenerate limit in App.~\ref{app:PSdegenerate}.
The volume emissivity of $U(1)_{\rm B-L}$ gauge bosons, which is dominated by the $J$-type matrix elements, reads as follows
\be
\begin{split}
\left. Q_{\rm Brem}^{n\text{-}p}\right|_{\rm deg}   \simeq & \, \frac{11}{270\pi^3}e^{\prime 2}  \left(\frac{f_\pi}{m_\pi}\right)^4 m_n^* m_p^* \,  p_{F_p}^3 T^4  J\left(y_p,x_{np}\right)\, \\
= & \, 4.7 \times 10^{26}\,{\rm erg}\,{\rm cm}^{-3}\,{\rm s}^{-1} \left(\frac{e^\prime}{10^{-9}}\right)^{2}\left(\frac{m_n^*}{m_n}\right)\left(\frac{m_p^*}{m_p}\right)\left(\frac{\left|\vec{p}_{F_p}\right|}{1/{\rm fm}}\right)^3 J\left(y_p,x_{np}\right)\, ,
\end{split}
\ee
where  $m_{N}^*$ denotes the effective nucleon mass, $y_i = m_\pi/2p_{F_i}$, $x_{np} = p_{F_n}/p_{F_p}$, and
\be
\begin{split}
J\left(y_p,x_{np}\right)   = & \,  F_6\left(y_p\right) - 2 I\left(y_p,x_{np}\right)+  \\
& +\frac{\left(x_{np}+1\right)^3}{4} \left(F_4\left(\frac{2 y_p}{x_{np}+1}\right)- F_6\left(\frac{2y_p}{x_{np}+1}\right)\right) + \\
& -\frac{\left(x_{np}-1\right)^3}{4} \left(F_4\left(\frac{2y_p}{x_{np}-1}\right)- F_6\left(\frac{2y_p}{x_{np}-1}\right)\right) +  \\
& +\frac{\left(x_{np}+1\right)\left(x_{np}-1\right)^2}{4} \left(F_4\left(\frac{2y_p}{x_{np}+1}\right)- F_2\left(\frac{2y_p}{x_{np}+1}\right)\right) + \\
&  -\frac{\left(x_{np}+1\right)^2\left(x_{np}-1\right)}{4} \left(F_4\left(\frac{2y_p}{x_{np}-1}\right)- F_2\left(\frac{2y_p}{x_{np}-1}\right)\right) \, 
\end{split}
\ee
with the functions $F_i(x)$ provided in App.~\ref{app:PSdegenerate} and the $I(y,x)$ function written as
\bea
 \int_0^{2p_{F_p}} \frac{d  |\mathbf{k} |}{2p_{F_p}} \int_0^{2\pi} \frac{d\phi_l}{2\pi} \frac{\mathbf{k}^2\mathbf{l}^2}{\left(\mathbf{k}^2+m_\pi^2\right)\left(\mathbf{l}^2+m_\pi^2\right)} \mathbf{k}^2 & = &\left(2p_{F_p}\right)^2 I\left(y_p,x_{np}\right) \, . \label{eq:Ifunction}
\eea
Unlike supernovae, we realize that the absorption (or trapping) argument is not applicable to the case of young NS cooling.
In both cases of supernovae and young NS, the density profile is concentrated within the core, so $U(1)_{\rm B-L}$ gauge bosons are dominantly produced inside the core.
However, 
the profile of the proto-neutron star for supernovae  accommodates an extensive space outside of the core~\cite{Chang:2016ntp} (e.g., up to neutrino sphere or the `far' radius). 
This makes $U(1)_{\rm B-L}$ gauge bosons efficiently absorbed into the medium. On the other hand, the profile of young NS contains a thin crust with rather small amount of neutrons and even no free proton.
Due to this key distinction, even if the typical mean free path reaches (or is much less than) the size of the core, a net discharge of $U(1)_{\rm B-L}$ gauge bosons can be still effective at outermost of the core of young NS.

\subsection{Constraints on dark $U(1)_{\rm B-L}$ gauge bosons}
\label{sec:constraints}

\begin{figure}[t!]
\centering
\includegraphics[width=0.8\textwidth]{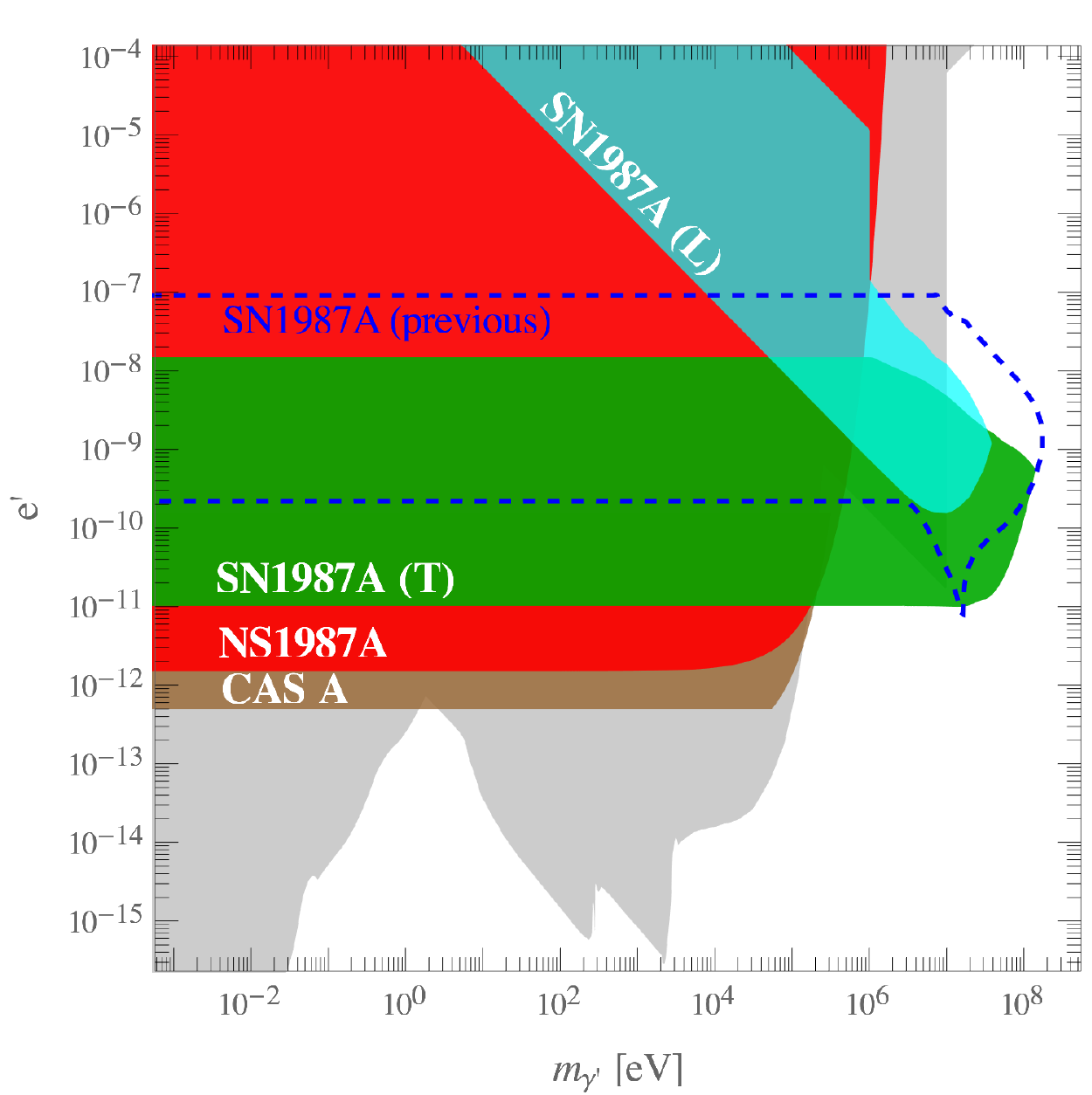}
\caption{\em The constraints on the $U(1)_{\rm B-L}$ gauge boson scenario.
  The green and cyan region are the revisited constraints on transverse (T) and longitudinal (L) components, respectively, from the observation of the neutrino flux for the first 10 second of SN1987A.
For comparison, the blue dashed line shows the previous constraint from SN1987A derived in Ref.~\cite{Knapen:2017xzo}.
The red region is excluded by the recent observations of NS1987A.
The gray region corresponds to the current constraints from the fifth force experiments~\cite{Murata:2014nra}, from big bang nucleosynthesis~\cite{Heeck:2014zfa,Knapen:2017xzo}, and from stellar cooling argument in the sun, HB stars, red giants~\cite{Hardy:2016kme,An:2014twa}, and Cas A~\cite{Hong:2020bxo}.}
\label{fig:bound}
\end{figure}

Let us explore how the parametric space is constrained by the stellar cooling argument on NS.
For SN1987A, we adopt the convenient criterion established by Raffelt~\cite{Raffelt:1996wa} that the net energy loss by a novel particle emission have to be smaller than $3\times 10^{52}\,{\rm erg}\,{\rm s}^{-1}$ to evade diminishing the duration of the observed supernovae neutrino signal; this is visualized with the gray line in Fig.~\ref{fig:Sn1987Acooling}.
We eventually find the constraint from SN1987A by following the discussion in the previous subsection to compute the gross emission rate, which incorporates a reabsorption into the medium.
The green and cyan regions in Fig.~\ref{fig:bound} show the parameter space excluded by SN1987A for each polarization of $U(1)_{\rm B-L}$ gauge bosons: transverse (T) and longitudinal (L).
The upper bounds on the coupling for transverse and longitudinal components are given by $e^\prime < 10^{-11} \, ({\rm T})$ and $e^\prime m_{\gamma^\prime}  < 7.4 \times10^{-10}\,{\rm MeV} \, ({\rm L})$ up to $m_{\gamma^\prime} < 20\,{\rm MeV} $. 
For $m_{\gamma^\prime} \gtrsim 20\,{\rm MeV}$, it is exponentially less stringent with respect to the mass.
We also have the lower bounds for transverse and longitudinal components
as that $e^\prime>1.5 \times 10^{-8} \,({\rm T})$ and $e^\prime m_{\gamma^\prime}>1.2 \times 10^{-5} \,{\rm MeV} \,({\rm L})$ are allowed for $m_{\gamma^\prime} < 2m_e \simeq 1\,{\rm MeV}$.
This lower bounds become weakened for the mass range above the threshold ($m_{\gamma'}=2m_e$) due to the additional absorption by the decay into electron-positron pair, which is in particular significant for longitudinal components.

 For comparison, we report the previous SN1987A bound in Ref.~\cite{Knapen:2017xzo} \footnote{There is  the recent study on the SN1987A bound via scatterings with abundant muons inside the core of the proto-neutron star~\cite{Croon:2020lrf}. In particular, the inclusion of neutrino-pair coalescence $\nu\bar{\nu} \rightarrow \gamma^\prime$ gives a difference at a large mass region for $m_{\gamma^\prime}>20\,{\rm MeV}$.} with the dashed blue line in Fig.~\ref{fig:bound}, which is an order of magnitude less stringent than our result.
The result in Ref.~\cite{Knapen:2017xzo} is based on the rescaled constraints on the dark photon scenario in Ref.~\cite{Chang:2016ntp} and the $U(1)_{\rm B}$ scenario in Ref.~\cite{Rrapaj:2015wgs}.
More specifically, the flat boundary for masses $m_{\gamma^\prime} < 3\,{\rm MeV}$ originates from the limit on $U(1)_{\rm B}$ gauge bosons~\cite{Rrapaj:2015wgs}.
Note that the $U(1)_{\rm B}$ gauge boson coupling to protons as well as neutrons are not screened by the plasma effect according to the discussion around Eq.~\eqref{eq:effectiveInMedium}, hence  $\tilde{q}_{p}^\prime = \tilde{q}_{n}^\prime $. This leads to the cancelation of the leading order contributions of the nucleon-nucleon bremsstrahlung scatterings, and the nonzero contribution occurs from the next-leading order (i.e., $\mathcal{M}_{a(\alpha,2)}$ as the corresponding matrix elements).
This is the main reason why our revision in the SN1987A limit on $U(1)_{\rm B-L}$ gauge bosons exhibits an order of magnitude severe than the previous results at the low mass range of $m_{\gamma^\prime} < 3\,{\rm MeV}$.
Alternatively, Ref.~\cite{Knapen:2017xzo} takes the rescaled result of the dark photon scenario~\cite{Chang:2016ntp} for higher masses $m_{\gamma^\prime} > 3\,{\rm MeV}$ because it is stronger.
We find $\tilde{q}_{p}^\prime-\tilde{q}_{ n}^\prime = m_{\gamma^\prime}^2/(m_{\gamma^\prime}^2-\pi_{\rm T,L})$ in the dark photon case. This is still different from Eq.~\eqref{eq:EffCoupqp-qn}, so the dark photon bound cannot be explicitly used.
Furthermore, this bound has a peak at $\omega  \sim 15\,{\rm MeV} \sim \left[ \omega_{\rm pl} \right]_{r<r_c}$, which comes from a resonance at the photon mediator of the dark photon coupling to protons.
However, such a resonance width calculated in the soft radiation approximation (corresponding to the $J$-type matrix elements) is underestimated.
We verify that considering all the types of matrix elements makes an width broad enough to undermine a resonance effect on the $U(1)_{\rm B-L}$ luminosity, then the constraint on $U(1)_{\rm B-L}$ gauge bosons from SN1987A has no peak.

\begin{figure}[t!]
\centering
\includegraphics[width=0.75\textwidth]{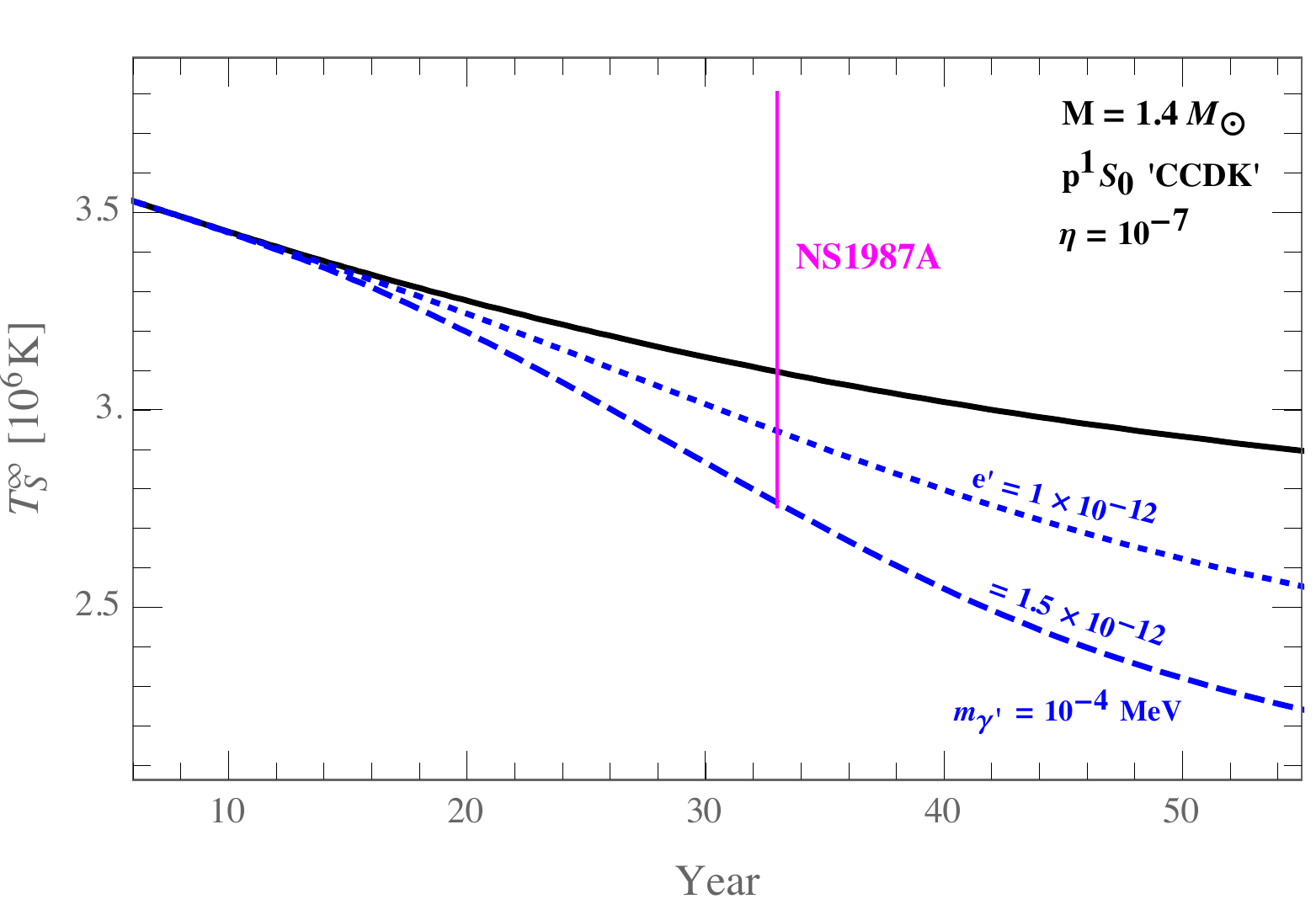}
\caption{\em  Cooling curves with the NS mass $M = 1.4 M_\odot$, the `CCDK'~\cite{Chen:1993bam} proton singlet pairing model and the amount of light elements in envelope given by $\eta = 10^{-7}$. The black solid line corresponds to the standard cooling scenario. The blue lines show the cooling curves in the $U(1)_{\rm B-L}$ gauge boson scenario for $m_{\gamma^\prime} =  10^{-4}\,{\rm MeV}$, and $e^\prime = 10^{-12}$ (dotted) and $1.5\times 10^{-12}$ (dashed), respectively.}
\label{fig:NS1987Acooling}
\end{figure}

For young NS cooling, we exploit the code `\texttt{NSCool}'~\cite{nscool} to carry out numerical simulations.
As in Ref.~\cite{Hong:2020bxo}, we employ the Akmal-Pandharipande-Ravenhall~\cite{Akmal:1998cf} equation of state with the non-magnetic carbon atmosphere (‘\texttt{APR-EOS-Cat.dat}’).
We assume a NS mass to be $1.4$ solar mass, which is within the reasonable range for a possible presence of the compact remnant of SN1987A as a neutron star called NS1987A~\cite{Page:2020gsx}.
The other important NS model parameter for NS cooling history is the amount of accreted light elements (such as H or He) in the  envelope~\cite{Wijngaarden:2019tht,Shternin:2021fpt,Page:2020gsx}, which controls the thermal conductivity  and hence the relation between the internal temperature and the surface temperature.
This amount is quantified by $\eta = g_{14}\Delta M /M_{\rm NS}$~\cite{Potekhin:1997mn}, where $g_{14}$ denotes the surface gravity in units of $10^{14}\,{\rm cm}\,{\rm s}^{-2}$ and $M_{\rm NS}$ and $\Delta M$ are the NS mass and the total mass of light elements, respectively.
In order to match the observationally inferred luminosity of NS1987A~\cite{Page:2020gsx,Cigan:2019shp} with the standard cooling scenario~\cite{Page:2004fy,Yakovlev:2004iq,Page:2009fu}, a rather large $\eta \gtrsim 10^{-8}$ is demanded to impose a relatively larger surface temperature with respect to the same core temperature~\cite{Hong:2020bxo}.
In Fig.~\ref{fig:NS1987Acooling}, the magenta line corresponds to the gravitationally red-shifted surface temperature of NS1987A from such x-ray observations.
With the choice of $\eta = 10^{-7}$, the solid black line describes the cooling curve in the standard cooling scenario and it is well within the expected range.
If the $U(1)_{\rm B-L}$ gauge boson couplings are strong enough as $e^\prime \gtrsim \mathcal{O}\left(10^{-12}\right)$, the energy loss by $U(1)_{\rm B-L}$ gauge bosons becomes comparable to the neutrino emission then it affects the NS cooling significantly.
As a result, we find the constraint on $U(1)_{\rm B-L}$ gauge bosons from NS1987A that allows $e^\prime < 1.5\times 10^{-12}$ for $m_{\gamma^\prime} \leq T_{\rm NS1987A} = \mathcal{O}\left(10^{9}\right) \,{\rm K} = \mathcal{O}\left(0.1\right)\,{\rm MeV}$ shown as the red region in Fig.~\ref{fig:bound}.
We note that, as discussed in the previous subsection, there is no lower bound from NS1987A due to its profile well condensed into the core.

\section{Conclusions and Discussion}
\label{sec:Conclusions}


We have discussed the production of dark gauge bosons via nucleon-nucleon bremsstrahlung from NS with considering their effective couplings to hadrons concretely in the context of the ChPT and the thermal field theory framework.
The dark gauge boson scenario with the ${\rm B-L}$ charge assignment is of our interest because a significant net difference between the effective couplings to protons and neutrons, which is associated with the leading order process of $n$-$p$ bremsstrahlung, arises from the in-medium mixing effect.
  We have revisited the constraint on $U(1)_{\rm B-L}$ gauge bosons from SN1987A; in the case of transverse polarizations, $e^\prime < 10^{-11}$ for $m_{\gamma^\prime}<20\,{\rm MeV}$, which is an order of magnitude more stringent than the earlier derivation of Ref.~\cite{Knapen:2017xzo}, and for $m_{\gamma^\prime}<1\,{\rm MeV}$, $e^\prime >1.5 \times 10^{-8}$  is allowed; for longitudinal components, the allowed region is $e^\prime m_{\gamma^\prime} < 7.4 \times10^{-10}\,{\rm MeV}$ for $m_{\gamma^\prime}<20\,{\rm MeV}$ and $e^\prime m_{\gamma^\prime}>1.2 \times 10^{-5}\,{\rm MeV}$ for $m_{\gamma^\prime}<1\,{\rm MeV}$.
We also find that NS1987A as a NS remnant of SN1987A provides the another stringent constraint on $U(1)_{\rm B-L}$ gauge bosons as $e^\prime < 1.5\times 10^{-12}$ for $m_{\gamma^\prime} \leq T_{\rm NS1987A} = \mathcal{O}\left(0.1\right)\,{\rm MeV}$ without any lower bound.

In this paper, we retained all the matrix elements not only of the $J$-type, which is relevant for the soft radiation approximation~\cite{Low:1958sn,Nyman:1968jro,Heller:1968cry,Rrapaj:2015wgs}, but also of the other types such as $\sigma$-type, `internal', and `contact'.
If the energy of produced $U(1)_{\rm B-L}$ gauge bosons is typically much less than the non-relativistic center of mass kinetic energy of nucleons (i.e., $\left<\omega\right>\sim T \ll \left<E_{\rm cm}\right>\sim \left<\mathbf{k}^2/2m_N\right> \sim \left<\mathbf{l}^2/2m_N\right>$ as in the degenerate limit), then the soft radiation approximation is valid and the $J$-type matrix elements dominate the emission rate.
In other words, the contributions of the other type matrix elements could be comparable as long as the condition $\left<\omega\right>\sim \left<E_{\rm cm}\right>\sim T$ is fulfilled as in the classical limit.
We can easily figure this features out with the non-relativistic expansion of the spinors $u(p_i)$ of nucleons, which appear in the matrix elements by the following three ways
\bea
\bar{u}\left(p^\prime\right)\gamma^5 u\left(p\right) & \simeq & - \xi^\dagger_{p^\prime} \vec{\sigma}\xi_p \cdot \left(\vec{p}^\prime - \vec{p}\right) \, , \\
\bar{u}\left(p^\prime\right)\gamma^\mu\gamma^5 u\left(p\right) & \simeq & - \xi^\dagger_{p^\prime} \vec{\sigma}_i \xi_p  \ \delta^{\mu i} 2m_N \, , \\
\bar{u}\left(p^\prime\right)i \sigma^{\mu\nu}\gamma^5 u\left(p\right) & \simeq & - \xi^\dagger_{p^\prime} \vec{\sigma}_i \xi_p  \left(\delta^{\mu 0}\delta^{\nu i}-\delta^{\mu i}\delta^{\nu 0}\right) 2m_N \, 
\eea
with a numerical two-component spinor $\xi_p$.
Assuming the pion masses are negligible, we perform the comparison of the leading order $u$-channel matrix elements of the $J$-type with those of the other types
\bea
\frac{|\mathcal{M}_{u(\sigma,1)}|}{|\mathcal{M}_{u(J,1)}|} \sim \frac{|\mathcal{M}_{u({\rm int},1)}|}{|\mathcal{M}_{u(J,1)}|} \sim \frac{|\mathcal{M}_{u({\rm con},1)}|}{|\mathcal{M}_{u(J,1)}|} \sim \mathcal{O}\left(\frac{\omega}{\mathbf{l}^2/m_N}\right) \, 
\eea
for summing over the transverse polarizations.
Therefore, our calculation in the classical limit may give an $\mathcal{O}(1)$ discrepancy in the emission rate compared to the result with the soft radiation approximation~\cite{Chang:2016ntp,Rrapaj:2015wgs}.
We also provide the same comparison for summing over the longitudinal component
\bea
\frac{|\mathcal{M}_{u(\sigma,1)}|_{\rm L}}{|\mathcal{M}_{u(J,1)}|_{\rm L}} \sim \mathcal{O}\left(\frac{\omega}{\mathbf{l}^2/m_N}\right)\, , \,\,  \frac{|\mathcal{M}_{u({\rm int},1)}|_{\rm L}}{|\mathcal{M}_{u(\sigma ,1)}|_{\rm L}} \sim \frac{|\mathcal{M}_{u({\rm con},1)}|_{\rm L}}{|\mathcal{M}_{u(\sigma ,1)}|_{\rm L}} \sim \mathcal{O}\left(\frac{\left|\mathbf{l}\right|}{m_N}\right) \, .
\eea
In this case, the `internal' and `contact' bremsstrahlung process are at the next-leading order in terms of the nucleon velocity, thus they can be neglected.

As a final remark, we comment on the uncertainties of our numerical analysis.
 One of the major theoretical uncertainties is associated with the approximation of one-pion-exchange to simplify the nucleon-nucleon interactions.
There are several effects to improve our computation based on this approximation, such as a finite pion mass in the mediator, the two-pions exchange, effective nucleon masses, and many-body effect from multiple nucleon scatterings.
As in Ref.~\cite{Carenza:2019pxu}, these effects may reduce the dark gauge boson production rate by an order of magnitude with respect to the approximated calculation.
The other significant theoretical uncertainties are involved in modelling the NS profile.
For SN1987A, we employ the `fiducial' model~\cite{Chang:2016ntp}, which is conveniently written by the simple analytic function in Eq.~\eqref{eq:fiducialModel}.
The fiducial model possesses the generic characteristics of the proto-neutron star that a vast majority of its mass and thermodynamic free energy is concentrated in the core ($\rho_c \approx 2.8 \times 10^{14}\,{\rm g}\,{\rm cm}^{-3}$ and $T\approx 30\,{\rm MeV}$).
Hence the derivation of the luminosity of $U(1)_{\rm B-L}$ gauge bosons in the free-streaming regime yielding the lower bound would be robust as in Ref.~\cite{Chang:2016ntp} for the dark photon scenario.
In fact, the absorptions of $U(1)_{\rm B-L}$ gauge bosons rely on the profile, in particular when $m_{\gamma^\prime} >  2m_e$ that the decay channel to electron-positron pairs is turned on.
In this paper, we take the largest possible value of the far radius as $1000\,{\rm km}$ so that the upper bound would be conservative.
For cooling of NS1987A before the presence of a weak neutron triplet superfluidity, large uncertainties arise from the amount of light elements in the envelope (for the discussion on other parameters such as mass and pairing gap model of nucleon superfluidity, see Ref.~\cite{Hong:2020bxo}).
As long as $\eta > 10^{-8}$, cooling curves of the standard cooling scenario can be intersected within the inferred surface temperature range of NS1987A as shown in Fig~\ref{fig:NS1987Acooling} with $\eta= 10^{-7}$ for a fiducial value.
Even if we take into account very large and implausible $\eta$ like $10^{-3}$, the expected surface temperature at the age of NS1987A in the standard cooling rises by a few tens of percent.
As a result, the red region in Fig.~\ref{fig:bound} is robustly excluded.





\paragraph{Acknowledgments.} 
The authors thank L. M. Krauss for useful discussions.
 The authors specially thanks to the referee to point out several important issues, in particular the phenomenological difference of each polarization.
This work is supported by the research grants: ``The Dark Universe: A Synergic Multi-messenger Approach'' number 2017X7X85K under the program PRIN 2017 funded by the Ministero dell'Istruzione, Universit\`a e della Ricerca (MIUR); ``New Theoretical Tools for Axion Cosmology'' under the Supporting TAlent in ReSearch@University of Padova (STARS@UNIPD). SY also supported by Istituto Nazionale di Fisica Nucleare (INFN) through the Theoretical Astroparticle Physics (TAsP) project.


\appendix


\section{Squared matrix elements}
\label{app:SqMatrix}

In this appendix, we provide the square of the matrix elements given in Sec.~\ref{sec:MatrixElements}.
They are not averaged over spins, but the angle with the dark gauge boson momentum $\vec{p}_{\gamma^\prime}$, i.e. $\hat{\mathbf{k}}\cdot \hat{p}_{\gamma^\prime}$ and/or $\hat{\mathbf{l}}\cdot \hat{p}_{\gamma^\prime}$.
We perform the polarization sum for the longitudinal component individually (denoted with `$\rm L$' subscript) as well as for the overall including the transverse components.

\subsection*{leading order}

The squared matrix elements for the $J$-type are given by
\begin{eqnarray}
\left|\mathcal{M}_{t(J,1)} \right|^2 & = &  e^{\prime 2} \left(\tilde{q}_1 - \tilde{q}_2\right)^2 \left(\frac{f_{13}f_{24}}{m_\pi^2}\right)^2\frac{\mathbf{k}^4}{\left(\mathbf{k}^2+m_\pi^2\right)^2} \,  \frac{128}{3} \mathbf{k}^2 \frac{m_N^2}{\omega^2} \, ,\\
\left|\mathcal{M}_{u(J,1)} \right|^2 & = & e^{\prime 2} \left(\tilde{q}_1 - \tilde{q}_2\right)^2 \left(\frac{f_{14}f_{23}}{m_\pi^2}\right)^2\frac{\mathbf{l}^4}{\left(\mathbf{l}^2+m_\pi^2\right)^2} \,   \frac{128}{3} \mathbf{k}^2 \frac{m_N^2}{\omega^2} \, ,\\
\mathcal{M}_{t(J,1)}^\dagger  \mathcal{M}_{u(J,1)} + {\rm h.c.} & = &  e^{\prime 2} \left(\tilde{q}_1-\tilde{q}_2\right)^2 \left(\frac{f_{13}f_{24}f_{14}f_{23}}{m_\pi^4}\right) \nn \\
 && \qquad  \times\frac{\mathbf{k}^2\mathbf{l}^2 -2 \left(\mathbf{k}\cdot \mathbf{l}\right)^2}{\left(\mathbf{k}^2+m_\pi^2\right)\left(\mathbf{l}^2+m_\pi^2\right)} \,   \frac{128}{3} \mathbf{k}^2 \frac{m_N^2}{\omega^2}\, . 
\end{eqnarray}
The leading order $\sigma$-type appears only in the $u$-channel
\begin{eqnarray}
\left| \mathcal{M}_{u(\sigma,1)} \right|^2 = e^{\prime 2}\left(\tilde{q}_1-\tilde{q}_2\right)^2 \left(\frac{f_{14}f_{23}}{m_\pi^2}\right)^2\frac{\mathbf{l}^2}{\left(\mathbf{l}^2+m_\pi^2\right)^2} \,  384 m_N^4  \, .
\end{eqnarray}
The cross terms of the $J$ and $\sigma$-type read
\begin{eqnarray}
\mathcal{M}_{u(\sigma,1)}^\dagger \mathcal{M}_{t(J,1)}  + {\rm h.c.} & = &  e^{\prime 2}\left(\tilde{q}_1-\tilde{q}_2\right)^2\left(\frac{f_{13}f_{24}f_{14}f_{23}}{m_\pi^4}\right) \frac{\mathbf{k}^2 }{\left(\mathbf{k}^2+m_\pi^2\right)\left(\mathbf{l}^2+m_\pi^2\right)} \nn\\
 && \times  \frac{256}{3} \frac{m_N^3}{\omega} \left(\mathbf{k}\cdot\mathbf{l}\right) \, ,  \\
\mathcal{M}_{u(\sigma,1)}^\dagger  \mathcal{M}_{u(J,1)}  + {\rm h.c.} & = & -  e^{\prime 2}\left(\tilde{q}_1-\tilde{q}_2\right)^2\left(\frac{f_{14}f_{23}}{m_\pi^2}\right)^2 \frac{\mathbf{l}^2 }{\left(\mathbf{l}^2+m_\pi^2\right)^2}  \frac{512}{3} \frac{m_N^3}{\omega}\left(\mathbf{k}\cdot\mathbf{l}\right) \, .
\end{eqnarray}

As discussed in Sec.~\ref{sec:MatrixElements} and shown in Fig.~\ref{fig:NNbremIC}, there are the additional diagrams associated with the dark gauge boson couplings to charged gauge bosons.
For the internal bremsstrahlung of the first diagram in Fig.~\ref{fig:NNbremIC}, we find
\begin{eqnarray}
\left| \mathcal{M}_{u({\rm int},1)}\right|^2 & = & e^{\prime 2} \left(\tilde{q}_1 - \tilde{q}_2\right)^2  \left(\frac{f_{14}f_{23}}{m_\pi^2}\right)^2  \frac{\mathbf{l}^6}{\left(\mathbf{l}^2+m_\pi^2\right)^4} \,  256 m_N^4  \, .
\end{eqnarray}
Its cross terms with the $J$ and $\sigma$-type are given by
\begin{eqnarray}
\mathcal{M}_{u({\rm int},1)}^\dagger\mathcal{M}_{t(J,1)} + {\rm h.c.} & = & e^{\prime 2} \left(\tilde{q}_1 - \tilde{q}_2\right)^2  \left(\frac{f_{13}f_{24}f_{14}f_{23}}{m_\pi^4}\right) \frac{\mathbf{k}^2\mathbf{l}^2-2\left(\mathbf{k}\cdot\mathbf{l}\right)^2}{\left(\mathbf{k}^2+m_\pi^2\right)\left(\mathbf{l}^2+m_\pi^2\right)^2} \nn\\
&& \times 128\frac{m_N^3}{\omega} \left(\mathbf{k}\cdot\mathbf{l}\right) \, ,  \\
\mathcal{M}_{u({\rm int},1)}^\dagger\mathcal{M}_{u(J,1)} + {\rm h.c.} & = & e^{\prime 2} \left(\tilde{q}_1 - \tilde{q}_2\right)^2  \left(\frac{f_{14}f_{23}}{m_\pi^2}\right)^2 \frac{\mathbf{l}^4}{\left(\mathbf{l}^2+m_\pi^2\right)^3} \, 256\frac{m_N^3}{\omega} \left(\mathbf{k}\cdot\mathbf{l}\right)  \, , \\
\mathcal{M}_{u({\rm int},1)}^\dagger \mathcal{M}_{u(\sigma,1)} + {\rm h.c.} & = & - e^{\prime 2} \left(\tilde{q}_1 - \tilde{q}_2\right)^2  \left(\frac{f_{14}f_{23}}{m_\pi^2}\right)^2 \frac{\mathbf{l}^4}{\left(\mathbf{l}^2+m_\pi^2\right)^3} \,  512 m_N^4 \, .
\end{eqnarray}
The contact bremsstrahlung of the right two diagrams in Fig.~\ref{fig:NNbremIC} has
\begin{eqnarray}
\left| \mathcal{M}_{u({\rm con},1)}\right|^2 & = & e^{\prime 2} \left(\tilde{q}_1 - \tilde{q}_2\right)^2  \left(\frac{f_{14}f_{23}}{m_\pi^2}\right)^2  \frac{\mathbf{l}^2}{\left(\mathbf{l}^2+m_\pi^2\right)^2} \,  512 m_N^4  \, ,
\end{eqnarray}
and also its cross terms with the $J$ and $\sigma$-type
\begin{eqnarray}
\mathcal{M}_{u({\rm con},1)}^\dagger\mathcal{M}_{t(J,1)} + {\rm h.c.} & = & e^{\prime 2} \left(\tilde{q}_1 - \tilde{q}_2\right)^2  \left(\frac{f_{13}f_{24}f_{14}f_{23}}{m_\pi^2}\right) \frac{\mathbf{k}^2}{\left(\mathbf{k}^2+m_\pi^2\right)\left(\mathbf{l}^2+m_\pi^2\right)} \nn\\
&&\times 128\frac{m_N^3}{\omega} \left(\mathbf{k}\cdot\mathbf{l}\right)  \, ,  \\
\mathcal{M}_{u({\rm con},1)}^\dagger\mathcal{M}_{u(J,1)} + {\rm h.c.} & = & - e^{\prime 2} \left(\tilde{q}_1 - \tilde{q}_2\right)^2  \left(\frac{f_{14}f_{23}}{m_\pi^2}\right)^2 \frac{\mathbf{l}^2}{\left(\mathbf{l}^2+m_\pi^2\right)^2}  \, 256\frac{m_N^3}{\omega} \left(\mathbf{k}\cdot\mathbf{l}\right) \, , \quad \\
\mathcal{M}_{u({\rm con},1)}^\dagger\mathcal{M}_{u(\sigma,1)} + {\rm h.c.} & = & e^{\prime 2} \left(\tilde{q}_1 - \tilde{q}_2\right)^2  \left(\frac{f_{14}f_{23}}{m_\pi^2}\right)^2 \frac{\mathbf{l}^2}{\left(\mathbf{l}^2+m_\pi^2\right)^2} \,  1024 m_N^4 \, .
\end{eqnarray}
Finally, we find the cross term of the internal and contact bremsstrahlung
\begin{eqnarray}
\mathcal{M}_{u({\rm int},1)}^\dagger\mathcal{M}_{u({\rm con},1)} + {\rm h.c.} & = & - e^{\prime 2} \left(\tilde{q}_1 - \tilde{q}_2\right)^2  \left(\frac{f_{14}f_{23}}{m_\pi^2}\right)^2 \frac{\mathbf{l}^4}{\left(\mathbf{l}^2+m_\pi^2\right)^3} \,  512 m_N^4 \, .
\end{eqnarray}

We also get the results with the polarization sum only over the longitudinal component, which are useful for the case of considering the plasma effect.
For the $J$ and $\sigma$-type, the squared matrix elements read
\begin{eqnarray}
\left|\mathcal{M}_{t(J,1)} \right|_{\rm L}^2 & = &  \frac{1}{2} \frac{m_{\gamma^\prime}^2}{\vec{p}_{\gamma^\prime}^2} \left|\mathcal{M}_{t(J,1)} \right|^2 \, , \\
\left|\mathcal{M}_{u(J,1)} \right|_{\rm L}^2  & = &  \frac{1}{2} \frac{m_{\gamma^\prime}^2}{\vec{p}_{\gamma^\prime}^2} \left|\mathcal{M}_{u(J,1)} \right|^2 \, ,  \\
\left[ \mathcal{M}_{t(J,1)}^\dagger  \mathcal{M}_{u(J,1)} + {\rm h.c.} \right]_{\rm L} & = & \frac{1}{2} \frac{m_{\gamma^\prime}^2}{\vec{p}_{\gamma^\prime}^2} \left( \mathcal{M}_{t(J,1)}^\dagger  \mathcal{M}_{u(J,1)} + {\rm h.c.} \right)\, , \\
\left| \mathcal{M}_{u(\sigma,1)} \right|_{\rm L}^2 & = & e^{\prime 2}\left(\tilde{q}_1-\tilde{q}_2\right)^2 \frac{m_{\gamma^\prime}^2}{\vec{p}_{\gamma^\prime}^2}  \left(\frac{f_{14}f_{23}}{m_\pi^2}\right)^2\frac{\mathbf{l}^2}{\left(\mathbf{l}^2+m_\pi^2\right)^2} \,  \frac{512}{3} m_N^4  \, ,\\
\left[ \mathcal{M}_{u(\sigma,1)}^\dagger \mathcal{M}_{t(J,1)}  + {\rm h.c.} \right]_{\rm L} & = &  e^{\prime 2}\left(\tilde{q}_1-\tilde{q}_2\right)^2 \frac{m_{\gamma^\prime}^2}{\vec{p}_{\gamma^\prime}^2} \left(\frac{f_{13}f_{24}f_{14}f_{23}}{m_\pi^4}\right) \frac{\mathbf{k}^2 }{\left(\mathbf{k}^2+m_\pi^2\right)\left(\mathbf{l}^2+m_\pi^2\right)} \nn\\
 && \times  \frac{128}{3} \frac{m_N^3}{\omega} \left(\mathbf{k}\cdot\mathbf{l}\right) \, ,  \\
\left[ \mathcal{M}_{u(\sigma,1)}^\dagger  \mathcal{M}_{u(J,1)}  + {\rm h.c.} \right]_{\rm L} & = & -  e^{\prime 2}\left(\tilde{q}_1-\tilde{q}_2\right)^2 \frac{m_{\gamma^\prime}^2}{\vec{p}_{\gamma^\prime}^2} \left(\frac{f_{14}f_{23}}{m_\pi^2}\right)^2 \frac{\mathbf{l}^2 }{\left(\mathbf{l}^2+m_\pi^2\right)^2} \nn \\
&&\times \frac{256}{3} \frac{m_N^3}{\omega}\left(\mathbf{k}\cdot\mathbf{l}\right) \, .  \ \ 
\end{eqnarray}
The common factor of $m_{\gamma^\prime}^2/\vec{p}_{\gamma^\prime}^2$ originates from current conservation.
We point out that the temporal component of the matrix elements of the internal and contact bremsstrahlung is sub-leading order.

\subsection*{next-leading order}

Now let us derive the squared matrix elements of next-leading order terms.
Similar to the leading order case, there are the eight diagrams shown in Fig.~\ref{fig:NNbremIC}, leading to the $J$ and $\sigma$-type matrix elements.
Note that there is no diagram for the internal and contact bremsstrahlung at the next-leading order.

For the $J$-type, we have
\begin{eqnarray}
\left| \mathcal{M}_{t(J,2)} \right|^2 & = & e^{\prime 2} \left(\tilde{q}_1+\tilde{q}_2\right)^2 \left(\frac{f_{13}f_{24}}{m_\pi^2}\right)^2 \frac{\mathbf{k}^4}{\left(\mathbf{k}^2+m_\pi^2\right)^2} \nn \\
&&\times  \frac{16}{\omega^2} \left[ \frac{2}{5}\mathbf{k}^2\mathbf{l}^2 + \frac{2}{15}\left(\mathbf{k}\cdot\mathbf{l}\right)^2\right] \ , \qquad  \\
\left| \mathcal{M}_{u(J,2)} \right|^2 & = & e^{\prime 2} \left(\tilde{q}_1+\tilde{q}_2\right)^2 \left(\frac{f_{14}f_{23}}{m_\pi^2}\right)^2 \frac{\mathbf{l}^4}{\left(\mathbf{l}^2+m_\pi^2\right)^2} \nn \\
&&\times \frac{16}{\omega^2} \left[ \frac{2}{5}\mathbf{k}^2\mathbf{l}^2 + \frac{2}{15}\left(\mathbf{k}\cdot\mathbf{l}\right)^2\right] \, ,\\
 \mathcal{M}_{t(J,2)}^\dagger \mathcal{M}_{u(J,2)}  + {\rm h.c.} & = & e^{\prime 2} \left(\tilde{q}_1+\tilde{q}_2\right)^2 \left(\frac{f_{13}f_{24}f_{14}f_{23}}{m_\pi^4}\right)\frac{\mathbf{k}^2\mathbf{l}^2-2 \left(\mathbf{k}\cdot \mathbf{l}\right)^2}{\left(\mathbf{k}^2+m_\pi^2\right)\left(\mathbf{l}^2+m_\pi^2\right)} \nn\\
 && \times  \,  \frac{16}{\omega^2} \left[ \frac{2}{5}\mathbf{k}^2\mathbf{l}^2 + \frac{2}{15}\left(\mathbf{k}\cdot\mathbf{l}\right)^2\right]  \, . 
\end{eqnarray}
The squared matrix elements of the $\sigma$-type are given by
\begin{eqnarray}
\left| \mathcal{M}_{t(\sigma , 2)} \right|^2 & = & e^{\prime 2} \left(\frac{\tilde{q}_1^2+\tilde{q}_1\tilde{q}_2+\tilde{q}_2^2}{3}\right) \left(\frac{f_{13}f_{24}}{m_\pi^2}\right)^2\frac{\mathbf{k}^4}{\left(\mathbf{k}^2+m_\pi^2\right)^2} \,  128 m_N^2  \, , \\
\left| \mathcal{M}_{u(\sigma,2)} \right|^2 & = & e^{\prime 2}\left(\frac{\tilde{q}_1+\tilde{q}_2}{2}\right)^2 \left(\frac{f_{14}f_{23}}{m_\pi^2}\right)^2\frac{\mathbf{l}^4}{\left(\mathbf{l}^2+m_\pi^2\right)^2} \,  128 m_N^2  \, ,\\
\mathcal{M}_{t(\sigma, 2)}^\dagger  \mathcal{M}_{u(\sigma, 2)}  + {\rm h.c.} & =  & -  e^{\prime 2} \left(\frac{\tilde{q}_1+\tilde{q}_2}{2}\right)^2 \left(\frac{f_{13}f_{24}f_{14}f_{23}}{m_\pi^4}\right)\frac{\left|\mathbf{k}\cdot \mathbf{l}\right|^2}{\left(\mathbf{k}^2+m_\pi^2\right)\left(\mathbf{l}^2+m_\pi^2\right)} \nn\\
&&\times  \frac{512}{3} m_N^2 \, .
\end{eqnarray}
We find their cross terms as following
\begin{eqnarray}
\mathcal{M}_{t(\sigma,2)}^\dagger \mathcal{M}_{t(J,2)} + {\rm h.c.} & = & - e^{\prime 2}\left(\tilde{q}_1+\tilde{q}_2\right)^2\left(\frac{f_{13}f_{24}}{m_\pi^2}\right)^2 \frac{\mathbf{k}^4}{\left(\mathbf{k}^2+m_\pi^2\right)^2} \frac{256}{15} \frac{m_N}{\omega}\left(\mathbf{k}\cdot\mathbf{l}\right) \, ,\\
 \mathcal{M}_{u(\sigma,2)}^\dagger\mathcal{M}_{u(J,2)}  + {\rm h.c.} & = & - e^{\prime 2}\left(\tilde{q}_1+\tilde{q}_2\right)^2\left(\frac{f_{14}f_{23}}{m_\pi^2}\right)^2 \frac{\mathbf{l}^4}{\left(\mathbf{l}^2+m_\pi^2\right)^2} \frac{256}{15} \frac{m_N}{\omega}\left(\mathbf{k}\cdot\mathbf{l}\right)\, ,\\
 \mathcal{M}_{t(\sigma,2)}^\dagger  \mathcal{M}_{u(J,2)} + {\rm h.c.} & = &e^{\prime 2}\left(\tilde{q}_1+\tilde{q}_2\right)^2\left(\frac{f_{13}f_{24}f_{14}f_{23}}{m_\pi^4}\right) \frac{\mathbf{k}^2\mathbf{l}^2 +\left(\mathbf{k}\cdot\mathbf{l}\right)^2}{\left(\mathbf{k}^2+m_\pi^2\right)\left(\mathbf{l}^2+m_\pi^2\right)} \nn \\
&&\times  \frac{64}{15}\frac{m_N}{\omega} \left(\mathbf{k}\cdot\mathbf{l}\right)   \, ,   \\
\mathcal{M}_{u(\sigma,2)}^\dagger  \mathcal{M}_{t(J,2)} + {\rm h.c.} & = &e^{\prime 2}\left(\tilde{q}_1+\tilde{q}_2\right)^2\left(\frac{f_{13}f_{24}f_{14}f_{23}}{m_\pi^4}\right) \frac{\mathbf{k}^2\mathbf{l}^2 +\left(\mathbf{k}\cdot\mathbf{l}\right)^2}{\left(\mathbf{k}^2+m_\pi^2\right)\left(\mathbf{l}^2+m_\pi^2\right)} \nn \\
&&\times  \frac{64}{15}\frac{m_N}{\omega} \left(\mathbf{k}\cdot\mathbf{l}\right) \, .
\end{eqnarray}

As in the leading order terms, we also provide the squared matrix elements with the summation over the only longitudinal polarization
\begin{eqnarray}
\left| \mathcal{M}_{t(J,2)} \right|_{\rm L}^2 & = & \frac{2}{3}\left| \mathcal{M}_{t(J,2)} \right|^2 \, , \\
 \left| \mathcal{M}_{u(J,2)} \right|_{\rm L}^2  & = &  \frac{2}{3}\left| \mathcal{M}_{u(J,2)} \right|^2 \, ,\\
\left[\mathcal{M}_{t(J,2)}^\dagger \mathcal{M}_{u(J,2)}  + {\rm h.c.}\right]_{\rm L} & = & \frac{2}{3}\left(\mathcal{M}_{t(J,2)}^\dagger \mathcal{M}_{u(J,2)}  + {\rm h.c.} \right) \, , \\
\left| \mathcal{M}_{t(\sigma , 2)} \right|_{\rm L}^2 & = & e^{\prime 2} \left(\frac{\tilde{q}_1^2}{3}+\frac{2\tilde{q}_1\tilde{q}_2}{5}+\frac{\tilde{q}_2^2}{3}\right) \left(\frac{f_{13}f_{24}}{m_\pi^2}\right)^2\frac{\mathbf{k}^4}{\left(\mathbf{k}^2+m_\pi^2\right)^2} \,  64 m_N^2  \, , \qquad  \\
\left| \mathcal{M}_{u(\sigma,2)} \right|_{\rm L}^2 & = & e^{\prime 2}\left(\frac{\tilde{q}_1+\tilde{q}_2}{2}\right)^2 \left(\frac{f_{14}f_{23}}{m_\pi^2}\right)^2\frac{\mathbf{l}^4}{\left(\mathbf{l}^2+m_\pi^2\right)^2} \,  \frac{1024}{15} m_N^2  \, ,\\
\left[ \mathcal{M}_{t(\sigma, 2)}^\dagger  \mathcal{M}_{u(\sigma, 2)}  + {\rm h.c.} \right]_{\rm L}& =  & - e^{\prime 2} \left(\frac{\tilde{q}_1+\tilde{q}_2}{2}\right)^2 \left(\frac{f_{13}f_{24}f_{14}f_{23}}{m_\pi^4}\right)\frac{\left|\mathbf{k}\cdot \mathbf{l}\right|^2}{\left(\mathbf{k}^2+m_\pi^2\right)\left(\mathbf{l}^2+m_\pi^2\right)} \nn\\
&&\times  \frac{256}{3} m_N^2 \, ,\\
\left[\mathcal{M}_{t(\sigma,2)}^\dagger \mathcal{M}_{t(J,2)} + {\rm h.c.} \right]_{\rm L}& = & - e^{\prime 2}\left(\tilde{q}_1+\tilde{q}_2\right)^2\left(\frac{f_{13}f_{24}}{m_\pi^2}\right)^2 \frac{\mathbf{k}^4}{\left(\mathbf{k}^2+m_\pi^2\right)^2} \frac{128}{5} \frac{m_N}{\omega}\left(\mathbf{k}\cdot\mathbf{l}\right) \, ,\\
\left[\mathcal{M}_{u(\sigma,2)}^\dagger\mathcal{M}_{u(J,2)}  + {\rm h.c.} \right]_{\rm L} & = & - e^{\prime 2}\left(\tilde{q}_1+\tilde{q}_2\right)^2\left(\frac{f_{14}f_{23}}{m_\pi^2}\right)^2 \frac{\mathbf{l}^4}{\left(\mathbf{l}^2+m_\pi^2\right)^2} \frac{128}{5} \frac{m_N}{\omega}\left(\mathbf{k}\cdot\mathbf{l}\right)\, ,\\
\left[ \mathcal{M}_{t(\sigma,2)}^\dagger  \mathcal{M}_{u(J,2)} + {\rm h.c.} \right]_{\rm L} & = & - e^{\prime 2}\left(\tilde{q}_1+\tilde{q}_2\right)^2\left(\frac{f_{13}f_{24}f_{14}f_{23}}{m_\pi^4}\right) \frac{\mathbf{k}^2\mathbf{l}^2 - 4 \left(\mathbf{k}\cdot\mathbf{l}\right)^2}{\left(\mathbf{k}^2+m_\pi^2\right)\left(\mathbf{l}^2+m_\pi^2\right)} \nn \\
&&\times  \frac{64}{15}\frac{m_N}{\omega} \left(\mathbf{k}\cdot\mathbf{l}\right)   \, ,   \\
\left[\mathcal{M}_{u(\sigma,2)}^\dagger  \mathcal{M}_{t(J,2)} + {\rm h.c.} \right]_{\rm L}& = & - e^{\prime 2}\left(\tilde{q}_1+\tilde{q}_2\right)^2\left(\frac{f_{13}f_{24}f_{14}f_{23}}{m_\pi^4}\right) \frac{\mathbf{k}^2\mathbf{l}^2 - 4\left(\mathbf{k}\cdot\mathbf{l}\right)^2}{\left(\mathbf{k}^2+m_\pi^2\right)\left(\mathbf{l}^2+m_\pi^2\right)} \nn \\
&&\times  \frac{64}{15}\frac{m_N}{\omega} \left(\mathbf{k}\cdot\mathbf{l}\right) \, .
\end{eqnarray}


\section{Phase space integral}
\label{app:PhaseSpace}

\subsection{classical limit}
\label{app:PSclassical}

In the classical limit where nucleons are non-relativistic and non-degenerate, we approximate their distribution function as $f_i  \simeq \left(n_N/2\right)\left(2\pi/m_N T\right)^{3/2} \exp\left[-|\vec{p}_i|^2/2m_N T\right]$ and $\left(1-f_3\right) \approx \left(1-f_4\right) \approx 1$.
We can neglect the spatial momentum of dark gauge bosons $\vec{p}_{\gamma^\prime}$ in the energy-momentum conservations law.
In the center-of-momentum coordinate, the momentums of nucleons in bremsstrahlung read
\begin{eqnarray}
\vec{p}_1 =\vec{P} + \vec{p}_i \, , \qquad \vec{p}_2 =\vec{P} - \vec{p}_i \, , \qquad \vec{p}_3 =\vec{P} + \vec{p}_f \, , \qquad \vec{p}_4 =\vec{P} - \vec{p}_f  \, .
\end{eqnarray}
Then we change the coordinate as follows
\bea
d\vec{p}_1 d\vec{p}_2 d\vec{p}_3 \, \Rightarrow \, 8 d \vec{p}_i d \vec{p}_f d \vec{P} \, .
\eea

We define the dimensionless variables
\bea
u = \frac{\vec{p}_i^2}{m_N T}\, , \,\, v = \frac{\vec{p}_f^2}{m_N T}\, ,\,\, x= \frac{\omega}{T}\, ,\,\, y = \frac{m_\pi^2}{m_N T}\, ,\,\, z= \cos\theta_{ij} = \frac{\vec{p}_i\cdot\vec{p}_f }{\left|\vec{p}_i\right|\left|\vec{p}_f\right|}\, ,\,\, q=\frac{m_{\gamma^\prime}}{T} \,  .
\label{eq:NDdimensionlessparameter}
\eea
The kinematic parameters, which often appear in the scattering amplitudes, are written in terms of those variables
\bea
\mathbf{k}^2 & = & m_N T \left(u+v - 2z \sqrt{uv}\right) \, , \\
\mathbf{l}^2  & = & m_N T \left(u+v + 2z \sqrt{uv}\right) \\
\left|\mathbf{k}\cdot\mathbf{l}\right|^2 & = & \left(m_N T\right)^2 \left(u-v\right)^2 \, .
\eea
Since the square of the matrix elements is independent of $\vec{P}$ due to the Lorentz invariance, we obtain
\bea
&& \left[ \prod_{i=1}^4 \int \frac{d \vec{p}_i}{2E_i(2\pi)^3} \right] f_1 f_2 \left(2\pi\right)^4\delta^{(4)} \left(p_1 + p_2 - p_3 - p_4 \mp p_{\gamma^\prime}\right)\nn\\
& = & \frac{n_N^2}{2^8 \pi^5 m_N^7 T^4}\int d \vec{p}_i d \vec{p}_f d \vec{P} \exp\left[-u\right]\exp\left[-\frac{\vec{P}^2}{m_N T}\right] \delta\left(u-v \mp x\right) \nn\\
& = & \frac{n_N^2 T^{1/2}}{2^7 \pi^{3/2} m_N^{5/2}} \int_0^\infty du \sqrt{u}e^{-u}\int_0^\infty dv \sqrt{v} \int_{-1}^{1} dz \, \delta\left(u-v \mp x\right)  \, .
\eea
In the first relation, the delta function for the spatial part is used to erase the $\vec{p}_4$ integration.
The sign of $p_{\gamma^\prime}$ in the delta function becomes $(-)$ or $(+)$ when a dark gauge boson is produced from or absorbed into the medium, respectively.
For production cases, the $\vec{p}_{\gamma^\prime}$ integration is included and given by
\bea
\int \frac{d \vec{p}_{\gamma^\prime}}{2\omega(2\pi)^3}  \omega \times \omega^{2n} = \frac{T^{3+2n}}{4\pi^2} \int_q^\infty dx x^{1+2n} \sqrt{x^2- q^2} \, ,
\label{eq:NDdarkgaugeintegral}
\eea
where the $\omega^{2n}$ factor comes from the sqaured matrix elements.

\subsection{degenerate limit}
\label{app:PSdegenerate}

For a non-relativistic and (strongly) degenerate Fermi gas, the momentum space below the Fermi momentum $p_{F_i}$ is exclusively filled.
In other words, only the narrow momentum shell near $p_{F_i}$, the width of which is controlled by the temperature fluctuation, contributes to nucleon scatterings such as bremsstrahlung.
Thus one can easily find the kinematics
\bea
\mathbf{k} \cdot  \mathbf{l}  & \approx & 0 \qquad \Big(\text{i.e.}\,\,\,\, \mathbf{k} \perp  \mathbf{l} \Big) \\
\mathbf{k} \times \mathbf{l} & \approx & \left| \mathbf{k}\right|\left| \mathbf{l}\right| \\
\mathbf{k}^2 + \mathbf{l}^2  & \approx &  \left| \vec{p}_1 - \vec{p}_2\right|^2 \approx  \left| \vec{p}_3 - \vec{p}_4\right|^2 \, .
\eea
Since $\mathbf{k}$ and $\mathbf{l}$ are nearly orthogonal, one can use them as the components of a vector in the multipole expansion~\cite{Nyman:1968jro,Rrapaj:2015wgs} of scattering amplitudes.
The differential of a spatial momentum for a degenerate case can be rewritten as
\bea
d \vec{p}_i \simeq \left( m_N p_{F_i} T \right) d x_i \, d\Omega_i \, ,
\eea
where $x_i = (E_i - \mu_{F,i})/T$.
We separate the phase space integral for nucleons' momentum out into the temporal part for $x_i$ and the angular part for $\Omega_i$ as in Ref.~\cite{Iwamoto:1992jp}.

\subsubsection*{temporal part}

One can extract the energy part of the phase space integral for nucleons in terms of $x_i$ parameters as following
\bea
\left[ \prod_{i=1}^4 \int d \vec{p}_i \right]_{E} & \rightarrow & \left[  \prod_{i=1}^{4} \int^{\infty}_{-\mu_{F,i}/T} d x_i\right]  f_1 f_2 \left(1-f_3\right) \left(1-f_4\right) \delta\left(T x_1 + T x_2 - T x_3 - T x_4 \mp \omega\right) \nn \\
& \approx & T^{-1}  \left[  \prod_{i=1}^{4}  \int^{\infty}_{-\infty} d x_i f_i\right] \delta\left(\sum_i x_i \mp y_a\right) \nn \\
& = & \frac{T^{-1}}{6}   \frac{y_a\left(y_a^2 + 4\pi^2\right)}{e^{\pm y_a}\mp 1} \, ,
\eea
where $y_a = \omega/T$, the sign $\mp$ of $\omega$ for production and absorption, respectively.
At the second line, we set all the $\mu_{F_i}/T \gg 1$ as infinity.
The $p_{\gamma^\prime}$ integration for production is straightforward
\bea
\left(\left[ \prod_{i=1}^4 \int d \vec{p}_i \right] \int \frac{d \vec{p}_{\gamma^\prime}}{\omega} \, \omega \times \omega^{2n} \right)_{E} & \rightarrow & \frac{T^{2+2n}}{6} \int_0^\infty dy_a \,  \frac{y_a^{3+2n}\left(y_a^2 + 4\pi^2\right)}{e^{y_a} - 1} \nn \\
& = & \left\{ 
\begin{tabular}{ccl}
$\frac{11 \pi^4}{90}$ & $\cdots$ & $n=-1$ \\
$\frac{62 \pi^6 T^2}{945}$ & $\cdots$ & $n=0$ 
\end{tabular}
\right.
\eea
with $\omega^n$ originating in the squared matrix elements.

\subsubsection*{angular part}

The angular part of the phase space integration can be rewritten in the two different ways with respect to remaining arguments of the integral.
The integration involving $\mathbf{k}$ is
\bea
&&\left(\prod_{i=1}^4 \int d\Omega_i\right) \delta^{(3)}\left(\vec{p}_3 + \vec{p}_4 - \vec{p}_1 - \vec{p}_2\right) \nn \\
& \simeq & \frac{8\pi^2}{p_1^2 p_2^2} \int_0^{2\,{\rm min}\left[\left|\vec{p}_1\right|,\left|\vec{p}_2\right|\right]} d  |\mathbf{k} | \int_0^{2\pi} d\phi_l 
\eea
with the redefinition of 
\bea
\mathbf{l}^2 = \left(|\vec{p}_1|^2 -\frac{\mathbf{k}^2}{4}\right)+\left(|\vec{p}_2|^2 -\frac{\mathbf{k}^2}{4}\right) - 2 \sqrt{|\vec{p}_1|^2 -\frac{\mathbf{k}^2}{4}}\sqrt{|\vec{p}_2|^2 -\frac{\mathbf{k}^2}{4}} \cos\phi_l \, .
\eea
Here, $\phi_l$ indicates the azimuthal angle of $\vec{p}_2$ (or $\vec{p}_4$) measured from $\mathbf{k}$.
Similarly, we could conduct the angular integration in terms of $\mathbf{l}$
\bea
&&\left(\prod_{i=1}^4 \int d\Omega_i\right) \delta^{(3)}\left(\vec{p}_3 + \vec{p}_4 - \vec{p}_1 - \vec{p}_2\right) \nn \\
&\simeq & \frac{8\pi^2}{p_1^2p_2^2} \int_{\left|p_1-p_2\right|}^{p_1+p_2} d  |\mathbf{l} | \int_0^{2\pi} d\phi_k
\eea
with the redefinition of 
\bea
\mathbf{k}^2 = \frac{\left[\left(|\vec{p}_1|+|\vec{p}_2|\right)^2-\mathbf{l}^2\right]\left[\mathbf{l}^2-\left(|\vec{p}_1|-|\vec{p}_2|\right)^2\right]}{\mathbf{l}^2}\frac{(1-\cos\phi_k)}{2} \, .
\eea

Angular integrations are mostly written as a combination of functions of
\bea
F_i \left(\tau\right) & = \int_0^1 dx \frac{x^i}{\left(x^2+\tau^2\right)^2} \, .
\eea 
We provide the analytic expressions for some $F_i$
\bea
F_2\left(\tau\right) & = & \frac{1}{2}\left( \frac{\arctan \left[1/\tau\right]}{\tau}- \frac{1}{2\left(\tau^2 +1\right)}\right) \, ,\\
F_4\left( \tau \right) & = & 1 - \frac{\tau^2}{2\left(\tau^2 +1\right)} - \frac{3}{2} \tau \arctan \left[1/\tau\right] \, ,\\
F_6\left(\tau\right) & = & \frac{1}{6}\left(5 - 15\tau^2 - \frac{3}{\tau^2+1} + 15 \tau^3 \arctan \left[1/\tau\right]\right) \, ,\\
F_8\left(\tau\right) & = & \frac{1}{5}-\frac{2\tau^2}{3}+3 \tau^4 +\frac{\tau^6}{2\tau^2 + 2} - \frac{7}{2}\tau^5 \arctan\left[1/\tau\right] \, .
\eea

\bibliographystyle{JHEP}
\bibliography{NNbremDP}

\end{document}